\documentclass[a4paper,11pt]{article}
\pdfoutput=1 

\usepackage{jheppub} 
\usepackage{mathtools}
\usepackage{bm}
\usepackage{amsmath,amssymb,amsthm,amscd,graphicx,simpler-wick}
\usepackage{psfrag}
\input epsf.sty
\usepackage{xparse}
\usepackage{tikz}
\usetikzlibrary{tikzmark}
\usepackage{tikz-cd}
\usepackage{needspace}
\usepackage{subcaption}
\usepackage{hyperref}

\usepackage{silence}
\usepackage[missing=]{gitinfo2}

\usepackage{varioref}
\usepackage{microtype}

\tikzset{snake it/.style={decorate, decoration=snake}}

\def\bbR{{\mathbb R}}
\def\bbC{{\mathbb C}}

\def\bbP{{\mathbb P}}

\newcommand{\psiplus}{{\psi_+}}
\newcommand{\psiminus}{{\psi_-}}

\newcommand{\tC}{{\widetilde C}}

\newcommand{\I}{{\mathrm i}}
\newcommand{\e}{{\mathrm e}}
\newcommand{\de}{\mathrm{d}}
\newcommand{\ri}{\text{i}}

\newcommand{\IP}[1]{\left\langle#1\right\rangle}

\newcommand{\eps}{\epsilon}

\newcommand{\cG}{{\mathcal G}}

\newcommand{\tJ}{{\widetilde{J}}}
\newcommand{\tT}{{\widetilde{T}}}

\newcommand{\cW}{{\mathcal W}}

\newcommand{\regular}{{\mathrm{reg}}}
\newcommand{\Conf}{{\mathrm{Conf}}}

\newcommand{\Heis}{{\mathrm{Heis}}}
\newcommand{\Vir}{{\mathrm{Vir}}}

\newcommand{\ren}{{\mathrm{ren}}}

\newcommand{\tot}{{\mathrm{tot}}}

\newcommand{\E}{\text{e}}

\newcommand{\rd}{\text{d}}

\newcommand{\tPsi}{{\widetilde \Psi}}

\newcommand{\even}{\text{even}}

\DeclareMathOperator{\sech}{sech}

\DeclareMathOperator{\disc}{disc}
\DeclareMathOperator{\nab}{{\mathcal F}}

\newcommand{\nop}[1]{{:}\, #1 {:} }

\usetikzlibrary{positioning,arrows,patterns}
\usetikzlibrary{decorations.markings}
\usetikzlibrary{calc}
\usetikzlibrary{shapes}
\usetikzlibrary{topaths}
\usetikzlibrary{decorations}
\usetikzlibrary{decorations.pathmorphing,fit}
\usetikzlibrary{bending}
\usetikzlibrary{calc}

\tikzset{
  fermion/.style={draw=black, postaction={decorate},decoration={markings,mark=at position .55 with {\arrow{>}}}},
  vertex/.style={draw,shape=circle,fill=black,minimum size=3pt,inner sep=0pt},
  vertex2/.style={draw,shape=circle,fill=black,minimum size=1pt,inner sep=0pt},
  vertex3/.style={draw,shape=circle,fill=\psiplusColor,minimum size=3pt,inner sep=0pt},
  vertex4/.style={draw,shape=circle,fill=\psiminusColor,minimum size=3pt,inner sep=0pt}  
}

\tikzset{dot/.style={circle, fill, inner sep=1.5pt}}

\newcommand{\branchpointColor}{orange}
\newcommand{\wallColor}{black}
\newcommand{\psiplusColor}{purple!70}
\newcommand{\psiminusColor}{blue!70}
\newcommand{\vdashLineColor}{gray!60} 
\newcommand{\jpColor}{black}      
\newcommand{\tpColor}{black}       
\newcommand{\coveringCurveColor}{gray} 
\newcommand{\baseCurveColor}{gray}  
\newcommand{\leashColor}{green!55!black}

\newlength{\branchpointmarkersize}
\newcommand{\drawbranchpointmarker}[1]{
  \draw [line width=0.5mm, \branchpointColor] ($(#1)-(\branchpointmarkersize/2,\branchpointmarkersize/2)$) -- ($(#1)+(\branchpointmarkersize/2,\branchpointmarkersize/2)$);
  \draw [line width=0.5mm, \branchpointColor] ($(#1)-(\branchpointmarkersize/2,-\branchpointmarkersize/2)$) -- ($(#1)+(\branchpointmarkersize/2,-\branchpointmarkersize/2)$);
}

\tikzset{
    psipluscontour/.style={thick,\psiplusColor},
    psiminuscontour/.style={thick,\psiminusColor},
    leash/.style={dashed,\leashColor}
  }

\title{\boldmath Nonabelianization map for the $W_{3,c=2}$ conformal blocks. Part I}
\author[1]{Qianyu Hao}
\preprint{MITP-26-067}
\affiliation[1]{PRISMA++ Cluster of Excellence \& Mainz Institute for Theoretical Physics\\
                 Johannes Guttenberg-Universit\"at\\
                 Staudinger Weg 7, 55128 Mainz, Germany}
\emailAdd{qiahao@uni-mainz.de}

\makeatletter
\gdef\@fpheader{\bigskip}
\makeatother

\begin{document}

\makeatletter
\renewcommand{\sectionautorefname}{\S\@gobble}
\renewcommand{\subsectionautorefname}{\S\@gobble}
\renewcommand{\subsubsectionautorefname}{\S\@gobble}
\makeatother

\abstract{
This work is inspired by the nonabelianization construction of Virasoro conformal blocks. We start the process of defining the nonabelianization map which expresses $c=2$ $W_3$-algebra conformal blocks on a Riemann surface $C$ from Heisenberg blocks on a branched triple cover  $\tC$ of $C$. This generalized nonabelianization map uses the datum of a $\mathfrak{gl}_3$ spectral network on $C$ and we find that this map also depends on the soliton degeneracies corresponding to the spectral network even though the soliton data didn't show up explicitly in the Virasoro case. 
This nonabelianization construction defines new formulas
for $W_3$-algebra conformal blocks on $C$ in terms of fermion correlation functions determined by the Heisenberg blocks
on $\tC$. We provide a check of the proposed construction associated to the more challenging $(123)$ type branch points which start to show up in three-fold covers. To be more precise, the simplest example that we focus on has only two of the $(123)$ type branch points and we show that in this example, our nonabelianization construction gives the desired two-point correlator and $\tau$-function. 
}

\maketitle

\flushbottom

\bibliographystyle{JHEP}

\section{Introduction}
Since the pioneer work \cite{Zamolodchikov:1985wn}, the two-dimensional conformal field theories with extended symmetries generated by higher spin currents whose vertex operator algebra (VOA) is $W_N$-algebra has been an active field to study. Foundational works for the $W_N$-algebra include the free field construction in terms of quantum Miura transformation \cite{Fateev:1987zh} as well as the quantum Drinfeld-Sokolov reduction \cite{Feigin:1990pn}. Compared to Virasoro algebra, the higher spin currents lead to difficulty in determining the correlation functions of descendants spanning the Verma module, for example, even the 3-point correlation functions can have infinite descendant constants. However it was found in $N=3$ that if one of the insertion is a semi-degenerate operator, the 3-point correlation functions can be determined up to a constant \cite{Bowcock:1993wq}; see also \cite{Coman:2017qgv} for a discrete family of basis. At $c=N-1$, by using the CFT/isomonodromy correspondence, \cite{Gavrylenko:2015wla} shows that the infinite series of constants for a type of conformal block with a degenerate field can be expressed in terms of the eigenvalues of Verlinde loop operators which are functions on the moduli space of flat connections. It is further claimed that this generalizes to all other conformal blocks. Even though closed-form expressions for generic $W$-algebra blocks are still missing, there has been solid progress on their definition and calculation. Here we only mention a non-exhaustive list of examples. One of the most fruitful resources is the generalized AGT correspondence \cite{Alday:2009aq}, which has been studied and applied extensively \cite{Wyllard:2009hg,MironovMorozov2010AGT,Kanno:2009ga}. This correspondence relating two-dimensional CFT and four-dimensional sypersymmetric gauge theory is very powerful since it provides instanton expansion expressions obtained from combinatorics for many $W$-algebra conformal blocks. Another impactful way which can be used to produce Fredholm determinant expressions of special $W$-algebra conformal blocks at $c=N-1$ is the TS/ST correspondence \cite{Grassi:2014zfa,Codesido:2015dia} (together with the AGT correspondence), used in for example \cite{Bonelli:2016idi,Gavrylenko:2023ewx}. 
More relevant to the present work are the exact conformal blocks for the generalized twist fields at $c=N-1$
\cite{Gavrylenko:2015cea,Gavrylenko:2016moe,Bershtein:2017lmg}. This approach used a branched free-field construction where a special type of $W_N$-conformal blocks on a Riemann surface $C$ is constructed using Heisenberg conformal blocks on a $N$-fold cover of $C$. However the resulting $W_N$-conformal blocks are limited to the ones with twist field insertions at branch points of the cover map.

In this paper, we develop a new way to construct the  $W_N$-algebra conformal blocks at $c=N-1$ which generalizes the nonabelianization map for conformal blocks in the $c=1$ Virasoro algebra (which is the same as $W_2$-algebra) case developed in \cite{Hao:2024vlg}, where spectral network plays a crucial role; see also \cite{Hollands:2026rhz} for more on spectral networks and Liouville conformal blocks. Our main goal is to construct a nonabelianization map to express the $W_N$-algebra conformal blocks in terms of the easier to compute Heisenberg blocks.  The nonabelianization map uses the branched free-field construction together with the data of spectral networks and soliton degeneracies. To be more precise, in this paper, we propose the nonabelianization map which applies to a generic rank $N$,
\begin{equation} \label{eq:intro-nab-map}
 \nab_\cW: \quad \Conf(\tC,\Heis) \quad \to \quad \Conf(C,W_{N,c=N-1} \otimes \Heis) \, ,
\end{equation}
where the algebra is enlarged by a copy of Heisenberg algebra (enlarging $W_{N}$ to $W_{N} \otimes \Heis$ is analogous to going from $\mathfrak{sl}_N$ to $\mathfrak{gl}_N$) and $\cW$ is a spectral network. For convenience, we sometimes refer to the $W_{N} \otimes \Heis$ case simply as the $W_{N}$ case; the intended meaning should be clear from the context.
 However, to write down explicit expressions, 
we focus on the simplest extended symmetry -- the $W_3$-algebra in this paper, i.e. $N=3$. The main result in this paper is to check that the proposed nonabelianization map is correct in a simple example which exhibits the new structure which doesn't exist in the Virasoro algebra case. 

To be more precise, in the free field construction, we consider a three-fold cover $\tC $ of a Riemann surface $C$, where the projection map is given by $\pi:\tC \xrightarrow{3:1}C$. Any point $p\in C$ can be lifted to three points $p^{(1)}$, $p^{(2)}$ and $p^{(3)}$ on $\tC $. Generators and primaries of the $W_3\otimes \Heis$-algebra on $C$ can be mapped to generators and primaries of the Heisenberg algebra on $\tC $ by matching the OPEs. As an example, the stress energy tensor $T(p)$ can be mapped to
\begin{equation} 
T(p) \rightsquigarrow \frac16 (\nop{(\tJ(p^{(1)}) - \tJ(p^{(2)}))^2}+\nop{(\tJ(p^{(2)}) - \tJ(p^{(3)}))^2} +\nop{(\tJ(p^{(3)}) - \tJ(p^{(1)}))^2} ).
\end{equation}
However, if we directly use this straightforward substitution, we will get a map with insertions of twist fields at the branch points of the covering $\pi$. In the $N=3$ case, the possible branch points are of two types. One is the type $(ij)$ which also appears in the $N=2$ Virasoro case. And the other is the type $(ijk)$ which is new to the higher ranks $N\geq 3$. In this paper, we focus on this new type of branch point and look at the simplest example where it could appear, where the projection map $\pi:\widetilde{C}\rightarrow C$ for both $\widetilde{C}$ and $C$ just $\mathbb{CP}_1$ is simply $z\mapsto x=z^3$.
There are only two branch points at $0$ and $\infty$, both of the type $(123)$. In this case the branch free field construction leads to a map 
$   \Conf(\tC,\Heis) \to \Conf\left(C,\Vir_{c=1}; W_{\frac{1}{9},0}(0)  W_{\frac{1}{9},0}(\infty) \right) ,
$ where the conformal weights of the twist fields can be calculated directly and they match with the result in \cite{Gavrylenko:2015cea}.

To achieve the goal of constructing a nonabelianization map for a generic $W_N$-algebra conformal blocks, we want a map \eqref{eq:intro-nab-map} without the twist fields. To do so, this present work proposes a generalization of \cite{Hao:2024vlg} which works out the nonabelianization map $ \nab_\cW$ for $c=1$ Virasoro conformal blocks without the twist fields. To be more precise, the  nonabelianization map for the $W_N$-algebra conformal blocks is achieved by inserting an extra $E(\cW)$ for any spectral network $\cW$ to the Heisenberg conformal blocks in the preimage of $ \nab_\cW$. In this work, we point out explicitly that $E(\cW)$ depends on the soliton data $\mu(-\widetilde{\gamma_{ji}}^{cb})$, meaning
\begin{equation} \label{eq:intro-EWreg-def}
E(\cW) =\exp \left[ -\frac{1}{2 \pi \I} \int_\cW\mu(-\widetilde{\gamma_{ji}}^{cb}) 
\tikz[remember picture, baseline]{
  \node[anchor=base, inner sep=0pt] (psiplus) {$\psi_+(q^{(+)})^{z^{(+)}}$};
}
\, 
\tikz[remember picture, baseline]{
  \node[anchor=base, inner sep=0pt] (psiminus) {$\psi_-(q^{(-)})^{z^{(-)}}$};
}
\, \de z \right]  \, ,
\end{equation}
which otherwise is almost the same as the one in \cite{Hao:2024vlg} if the $\mathfrak{gl}_2
$ spectral network $\cW$ is replaced by a $\mathfrak{gl}_N
$ spectral network.The soliton data are explained in \autoref{sec:virsol} and calculated for the example we study in \autoref{sec:solitonW3}. One might worry about this difference in the definition of $E(\cW)$ compared to  \cite{Hao:2024vlg}.
 But we show in \autoref{sec:matchsoln} that \cite{Hao:2024vlg} actually fixed a convention which is enough to suppress this dependence of the soliton data in the $N=2$ case so in fact there is no contradiction.

In principle, the cancellation of the twist fields and the insertion of \eqref{eq:intro-EWreg-def} is a local phenomenon. Thus in this paper, we check the nonabelianization map by the simplest example mentioned above where the projection map $\pi:\widetilde{C}\rightarrow C$ is given by $z\mapsto x=z^3$. The detailed calculation is included in \autoref{sec:tau}. This simple model contains only two important branch points of $(123)$ type which is new in the  examples with $W_3$-algebra. In this example, we calculated the normalized 2-fermion correlator $
F(z,w) = \frac{ \langle
\tikz[remember picture, baseline]{
  \node[anchor=base, inner sep=0pt] (psiplus1) {$\psi_+(v)$};
}
\,
\tikz[remember picture, baseline]{
  \node[anchor=base, inner sep=0pt] (psiminus1) {$\psi_-(w)$};
}
\, E_\ren(\cW) \rangle_\tPsi }{ \langle E_\ren(\cW) \rangle_\tPsi} \, 
$ and the $\tau$-function $\log \IP{E(\cW)}_{\tPsi}$, where $\tPsi\in\Conf(\tC,\Heis)$. They show that the nonabelianization construction works as expected. We further check that the conformal blocks with one Virasoro generator $T$ inserted is nonsingular at the branch points which provides evidence for the absence of the unwanted twist field.

In this paper, we focus on the most basic version of the nonabelianization of conformal blocks and provide some checks it in the simplest example. This leaves many open questions:
\begin{itemize}
\item It would be nice to also provide check for the higher spin current $\cW_3$. Also, the simplest example used in this paper is not very useful to check the nonabelization map with insertions of primaries, and leashes \cite{Hao:2024vlg} for higher genus surfaces.  More interesting examples and calculations are preferred to check our proposal. Moreover, the focus of this paper is to develop the essential part of the nonabelianization construction -- the insertion of the operator corresponding to a spectral network. However, it is still necessary to complete the story by including the structure of the space of conformal blocks, the discussion of Verlinde loop operators which are crucial when the space of conformal blocks are of infinite dimension, and the compatibilities.

\item We expect that the new construciton of $W_3$-algebra (or  $W_3\otimes \Heis$-algebra) conformal blocks would result in a continuous family of independent blocks on $C$ when Verlinde loop operators are diagonalized. An especially interesting example is the 3-punctured sphere where the result is still incomplete.

\item It would be very interesting to further generalize the nonabelianization map to the $q$-deformed $W_N$ vertex algebras. A straightforward attempt is to replace  spectral networks by exponential networks and use the corresponding soliton degeneracies.
\end{itemize}
\section{Vertex algebras and conformal blocks}
\subsection{Vertex algebras and primary fields}
The essential vertex algebra in this work is the $(W_3)$-algebra. A $(W_3)_c$-algebra at central charge $c$ is generated by a field $T$ and a primary field $W_3$ of it. 
It is important to specify local coordinate $z$ for fields. A field in a coordinate $z$ is represented by $T^z$ and $W_3^z$; however when the coordinate is clear from the context, we sometimes omit it in this paper.
At a central charge $c$, $T$-$T$,  $T$-$\cW_3$,  $\cW_3$-$\cW_3$ operator products are
\begin{align} 
\label{eq:virasoro-ope}
 T(p) T(q) = &\, \frac{c/2}{(z(p)-z(q))^4} + \frac{2 T(q)}{(z(p)-z(q))^2} + \frac{\partial_{z(q)} T(q)}{z(p)-z(q)} + \regular \, .
 \end{align}
 \begin{align}
 T(p) \cW_3(q) =&\, \frac{3 \cW_3(q) }{(z(p)-z(q))^2} +\frac{\partial_{z(q)} \cW(q)}{z(p)-z(q)} + \regular \, .
 \end{align}
 \begin{align}
\label{W3W3OPE}
\nonumber \cW_3(p) \cW_3(q) =&\, \frac{c/3 }{(z(p)-z(q))^6} + \frac{2 T(q)}{(z(p)-z(q))^4}  + \frac{\partial_{z(q)} T(q)}{(z(p)-z(q))^3}\\
  \nonumber&+\frac{1}{(z(p)-z(q))^2}\left(\frac{32}{22+5c}\Lambda(q)+\frac{3}{10}\partial^2_{z(q)}T(q)\right) \\
  &+\frac{1}{(z(p)-z(q))}\left(\frac{16}{22+5c}\partial_{z(q)}\Lambda(q)+\frac{1}{15}\partial^3_{z(q)}T(q)\right)+\regular \, ,
\end{align}
where $\Lambda(q)=\nop{TT}(q)-\frac{3}{10}\partial^2_{z(q)}T(q)$.

Under change of coordinates, $T$ transforms as
\begin{equation} \label{eq:virasoro-coordinate-change}
  T(p)^{z} = \left( \frac{\de w(p)}{\de z(p)} \right)^2 T(p)^{w} + \frac{c}{12} \{w,z\} \, ,
\end{equation}
where $\{\cdot,\cdot\}$ denotes the Schwarzian derivative:
$\{f(z),z\} = \frac{f'''(z)}{f'(z)} - \frac32 \left( \frac{f''(z)}{f'(z)} \right)^2$. And $\cW_3$ transforms as
\begin{align} \label{eq:W3-coordinate-change}
\cW_3(p)^z=(\frac{\rd w}{\rd z})^3\cW_3(p)^w.
\end{align}

The $W_3$-algebra primary fields $W_{h,w}$, for $h,w \in \bbC$, are defined by the conditions
\begin{equation} \label{eq:w3-primary-ope-T}
  T(p) W_{h,w}(q) = \frac{h W_{h,w}(q)}{(z(p)-z(q))^2} + \frac{\regular}{z(p) - z(q)} \, ,
\end{equation}

\begin{equation} \label{eq:w3-primary-ope-w3}
 \cW_3(p) W_{h,w}(q) = \frac{w W_{h,w}(q)}{(z(p)-z(q))^3} +\frac{W_{-1} W_{h,w}(q)}{(z(p)-z(q))^2} + \frac{W_{-2} W_{h,w}(q)}{z(p) - z(q)}+\regular \, .
\end{equation}
The constants $h$ and $w$ are the conformal weight and w-charge of the primary $W_{h,w}$. It is convenient to paramatrize them using the background charge $\bm{Q}$
\begin{equation}
c=2-12 \mathbf{Q}\cdot\mathbf{Q},\quad\quad \mathbf{Q}=(\frac1b-b)\bm{\rho},
\end{equation}
where $\bm{\rho}$ is the Weyl vector of $\mathbf{sl}_3$ and a charge vector $\bm{\alpha}$ as 
\begin{equation}
h_{\bm{\alpha}}=\frac12\bm{\alpha}\cdot(\bm{\alpha}-2\bm{Q}),
\end{equation}
\begin{equation}
w_{\bm{\alpha}}=\sqrt{\frac{48}{22+5c}}(\bm{\alpha}-\bm{Q})\cdot\bm{\omega}_1\times(\bm{\alpha}-\bm{Q})\cdot(\bm{\omega}_2-\bm{\omega}_1)\times(\bm{\alpha}-\bm{Q})\cdot(-\bm{\omega}_2),
\end{equation}
where $\bm{\omega}_i$ are weights of the Lie algebra\footnote{$\bm{\omega}_1\cdot\bm{\omega}_1=\bm{\omega}_2\cdot\bm{\omega}_2=\frac23$ and $\bm{\omega}_1\cdot\bm{\omega}_2=\frac13$ \cite{Dupic:2018pqr}.
}.

Introduce the parametrization $\bm{\alpha}=u\bm{\omega}_1+v\bm{\omega}_2$. In this paper, we focus on $c=2$, so\footnote{There are different choices of $u,v$ corresponding to the same $h,w$. They are all equivalent, so in this paper we will use an arbitrary one.}  
\begin{equation}\label{huv}
h_{u,v}=\frac13(u^2+uv+v^2),
\end{equation}
\begin{equation}\label{wuv}
w_{u,v}=\sqrt{\frac{3}{2}}(\frac{2 u^3}{27}+\frac{u^2 v}{9}-\frac{u v^2}{9}-\frac{2 v^3}{27}).
\end{equation}

Another important type of conformal blocks in this work is
\begin{equation} 
\Conf\left(C,W_{3,c=2} \otimes \Heis\right) \, ,
\end{equation}
where we the Heisenberg vertex algebra $\Heis$ is also used. The Heisenberg vertex algebra $\Heis$ is generated by one field $J$, with the OPE relation
\begin{equation} \label{eq:heisenberg-ope}
  J(p) J(q) = \frac{1}{(z(p)-z(q))^2} + \regular.
\end{equation}
Its primary fields $V_\alpha$, for $\alpha \in \bbC$, are defined by the condition
\begin{equation} \label{eq:heisenberg-primary-ope}
  J(p) V_\alpha(q) = \frac{\alpha V_\alpha(q)}{z(p) - z(q)} + \regular .
\end{equation}
Under change of coordinates, $J$ transforms as
\begin{equation} \label{eq:heisenberg-coordinate-change}
      J(p)^{z} = \left(\frac{\de w(p)}{\de z(p)}\right) J(p)^{w} \, . 
\end{equation}
The Heisenberg algebra contains a Virasoro algebra of central charge $1$, with generator 
$T^{\Heis}$ given by
\begin{equation} \label{eq:t-heis}
T^{\Heis}(p) = \frac12 \nop{J(p)^2} \, .
\end{equation}
Restricting attention to $T^\Heis$, the primary field $V_\alpha$ for $\Heis$
becomes a $\Vir_{c=1}$ primary $W_{\frac{\alpha^2}{2}}$. Indeed, $V_\alpha$'s conformal weight is given by $\frac{\alpha^2}{2}$.

\subsection{Conformal blocks}

To start with, we first look at the simplest vertex algebra $\Heis$. We define a complex vector space 
$\Conf(C,\Heis)$ as the space of Heisenberg conformal blocks on $C$. The definition of Heisenberg conformal blocks can be found in \cite{Hao:2024vlg} appendix A.1, which we reproduce here for convenience.

An element $\Psi \in \Conf(C,\Heis)$ represents a system of correlation
functions defined as follows. For every $n$, and any collection of patches 
$U_i$ on $C$ with local coordinate systems $z_i: U_i \hookrightarrow \bbC$,
we have a function
\begin{equation} \label{eq:heis-correlation-function}
  \IP{J( p_1 )^{z_1} \cdots J( p_n )^{z_n}}_\Psi : U_1 \times U_2 \times \cdots \times U_n \to \bbC \, .
\end{equation}

This collection of functions satisfies:
\begin{enumerate}
  \item Each $\IP{J( p_1 ) \cdots J( p_n )}_\Psi$ is meromorphic in $p_i$, with singularities only
  when some $p_i = p_j$.
  \item The collection is invariant under the symmetric group $S_n$, so e.g. 
  \begin{equation}
  \IP{J(p_1)^{z_1} J(p_2)^{z_2} \cdots}_\Psi = \IP{J(p_2)^{z_2} J(p_1)^{z_1} \cdots }_\Psi \, .  
  \end{equation}
    Here and below, $\cdots$ denotes an arbitrary product of insertions $J(p_i)^{z_i}$ which is the same on both sides of the equation. 
  \item When $p_1 \to p_2$ and they are in a single common coordinate system $U_1 = U_2 = U$ and $z_1 = z_2 = z$, the singularity of the $n$-point function as $p_1 \to p_2$ is determined
  by the $(n-2)$-point function, as
  \begin{equation} \label{eq:pole-condition-heisenberg-blocks}
    \IP{J(p_1)^z J(p_2)^z \cdots }_\Psi \ = \  \frac{1}{(z(p_1) - z(p_2))^2} \IP{\cdots}_\Psi + \regular.
  \end{equation}
More informally,
 \eqref{eq:pole-condition-heisenberg-blocks} says that the OPE relation \eqref{eq:heisenberg-ope} ``holds in correlation functions.''
  \item If $z$ and $z'$ are two local coordinate systems around $p$, then the correlation functions are related
  by
  \begin{equation} \label{eq:coordinate-transformation-heisenberg-blocks}
    \IP{J(p)^{z'} \cdots}_\Psi = \left(\frac{\de z(p)}{\de z'(p)}\right) \IP{J(p)^{z} \cdots}_\Psi \, .
  \end{equation}
\end{enumerate}

Note that all our conditions are linear over the complex numbers, 
so $\Conf(C,\Heis)$ is indeed a vector space, with the rule
\begin{equation}
  \IP{\cdots}_{a \Psi + b \Psi'} = a \IP{\cdots}_\Psi + b \IP{\cdots}_{\Psi'} \, .
\end{equation}

Similarly, given a Riemann surface $C$, we define a complex vector space 
$\Conf(C,W_3)$, the space of $W_3$ conformal blocks on $C$ in the following way:

An element $\Psi \in \Conf(C,W_3)$ is a system of correlation
functions. For every $n$, and any collection of patches 
$U_i$ on $C$ with local coordinate systems $z_i: U_i \hookrightarrow \bbC$,
there is a function
\begin{equation} \label{eq:heis-correlation-function}
  \IP{T( p_1 )^{z_1} \cdots T( p_n )^{z_n}\cW_3( q_1 )^{z_{n+1}} \cdots \cW_3( q_m )^{z_{n+m}}}_\Psi : U_1 \times U_2 \times \cdots \times U_{n+m} \to \bbC \, .
\end{equation}

This collection of functions has the following properties:
\begin{enumerate}
  \item Each $ \IP{T( p_1 )^{z_1} \cdots T( p_n )^{z_n}\cW_3( q_1 )^{z_{n+1}} \cdots \cW_3( q_m )^{z_{n+m}}}_\Psi $ is meromorphic in $p_i$, with singularities only
  when some $p_i = p_j$.
  \item The collection is invariant under the symmetric group $S_{n+m}$, so e.g. 
  \begin{equation}
  \IP{T(p_1)^{z_1} T(p_2)^{z_2} \cdots}_\Psi = \IP{T(p_2)^{z_2} T(p_1)^{z_1} \cdots }_\Psi \, .  
  \end{equation}
    \begin{equation}
  \IP{\cW_3(p_1)^{z_1} \cW_3(p_2)^{z_2} \cdots}_\Psi = \IP{\cW_3(p_2)^{z_2} \cW_3(p_1)^{z_1} \cdots }_\Psi \, .  
  \end{equation}
    \begin{equation}
  \IP{T(p_1)^{z_1} \cW_3(p_2)^{z_2} \cdots}_\Psi = \IP{\cW_3(p_2)^{z_2} T(p_1)^{z_1} \cdots }_\Psi \, .  
  \end{equation}
  \item If $U_1 = U_2 = U$ and $z_1 = z_2 = z$, then the singularity of the $n$-point function as $p_1 \to p_2$ is determined
  by the $(n-2)$-point function, as
  \begin{equation} \label{eq:pole-condition-virasoro-ope}
  \IP{T(p_1)^z T(p_2)^z\cdots} =\frac{c/2\IP{\cdots}_\Psi}{(z(p_1)-z(p_2))^4} + \frac{2 \IP{T(p_2)\cdots}_\Psi}{(z(p_1)-z(p_2))^2} + \frac{\IP{\partial_{z(p_2)} T(p_2)\cdots}_\Psi}{z(p_1)-z(p_2)} + \regular \, ,
\end{equation}
\begin{equation}\label{eq:pole-conditionTW3OPE}
  \IP{ T(p_1)^z \cW_3(p_2)^z} = \frac{3 \IP{\cW_3(p_2)\cdots}_\Psi }{(z(p_1)-z(p_2))^2} +\frac{\IP{\partial_{z(p_2)} \cW(p_2)\cdots}_\Psi}{z(p_1)-z(p_2)} ) + \regular \, ,
\end{equation}
\begin{equation}\label{eq:pole-conditionW3W3OPE}
\begin{aligned}
   \IP{ \cW_3(p_1)^z \cW_3(p_2)^z} &= (\frac{c/3 \IP{\cdots}_\Psi}{(z(p_1)-z(p_2))^6} + \frac{2 \IP{T(p_2)\cdots}_\Psi}{(z(p_1)-z(p_2))^4}  + \frac{\IP{\partial_{z(p_2)} T(p_2)\cdots}_\Psi}{(z(p_1)-z(p_2))^3}\\
  &+\frac{1}{(z(p_1)-z(p_2))^2}\left(\frac{32}{22+5c}\IP{\Lambda(p_2)\cdots}_\Psi+\frac{3}{10}\IP{\partial^2_{z(p_2)}T(p_2)\cdots}_\Psi\right) \\
  &+\frac{1}{(z(p_1)-z(p_2))}\left(\frac{16}{22+5c}\IP{\partial_{z(p_2)}\Lambda(p_2)\cdots}_\Psi+\frac{1}{15}\IP{\partial^3_{z(p_2)}T(p_2)\cdots}_\Psi\right))\\
  &+\regular \, .
\end{aligned}
\end{equation}

 In fact, \eqref{eq:pole-condition-virasoro-ope}-\eqref{eq:pole-conditionW3W3OPE} are determined by the OPE relation \eqref{eq:virasoro-ope}-\eqref{W3W3OPE}.
  \item If $z$ and $z'$ are two local coordinate systems around $p$, then the correlation functions are related
  by
  \begin{equation} \label{eq:coordinate-transformation-heisenberg-blocks}
    \IP{T(p)^{z'} \cdots}_\Psi = \left( \frac{\de z(p)}{\de z'(p)} \right)^2  \IP{T(p)^{z} \cdots}_\Psi + \frac{c}{12} \{w,z\}  \IP{\cdots}_\Psi \, ,
  \end{equation}
\begin{equation}
    \IP{\cW_3(p)^{z'}\cdots}_\Psi=(\frac{\rd z}{\rd z'})^3\IP{\cW_3(p)^z\cdots}_\Psi.
\end{equation}
\end{enumerate}

We further define a complex vector space for an enlarged algebra 
$\Conf(C,W_{3,c=2} \otimes \Heis)$ on $C$, where the poles on the diagonals and transformation laws are obtained as the combination of the ones of $\Conf(C,W_{3,c=2})$ and $\Conf(C,\Heis)$ while $J$-$T$ and $J$-$\cW_3$ operator products have no singularity.

In this work, we express the harder to compute $W_3$ conformal blocks in terms of the easier to deal with Heisenberg conformal blocks. In particular, we will do calculations in terms of Heisenberg conformal blocks, where $C = \bbC\bbP^1$. In this case,
$\Conf(C,\Heis)$ is $1$-dimensional\footnote{When $C$ is a compact surface of genus $g > 0$, $\Conf(C,\Heis)$ is infinite-dimensional.}.
Thus we can fix a block $\Psi$ by
choosing the zero-point function: we choose
\begin{equation}
\IP{1}_\Psi = 1 \, .  
\end{equation}
Note that the space $\Conf(C,W_{3,c=2})$ or $\Conf(C,W_{3,c=2} \otimes \Heis)$ is also $1$-dimensional in the case $C = \bbC \bbP^1$.

A more general definition of conformal blocks also include primaries in the correlation functions. Now fix points $q_1, \dots, q_k \in C$ 
and weights ${h_1,w_1}, \dots, {h_k,w_k} \in \bbC$ for $W_3$ primaries. Then we also define a vector space $\Conf(C,W_{3} ; W_{h_1,w_1}(q_1) \cdots W_{h_k,w_k}(q_k))$,
the space of $W_{3} $ conformal blocks on $C$ with primary fields $W_{h_j,w_j}(q_j) $ inserted.

The definition is almost the same as above and we make following changes.
The correlation functions are now denoted by 
\begin{align}
 \nonumber&\langle T( p_1 )^{z_1} \cdots T( p_n )^{z_n}\cW_3( p_{n+1} )^{z_{n+1}} \cdots \cW_3( p_{n+m} )^{z_{n+m}}W_{h_1,w_1}(q_1) \cdots W_{h_k,w_k}(q_k)\rangle_\Psi \\
 &: U_1 \times U_2 \times \cdots \times U_{n+m} \to \bbC \, .
\end{align}
(We emphasize that for now they are functions only of the $p_i$, not of the $q_j$ which are held fixed throughout the definition
of $\Conf(C,W_{3}; W_{h_1,w_1}(q_1) \cdots W_{h_k,w_k}(q_k))$.)
These correlation functions now also have poles at $p_i = q_j$ (as well as at $p_i = p_j$ as before).
The singularity of the $n$-point function as $p_1 \to q_j$ is determined by the $(n-1)$-point function, as

\begin{equation} 
      \IP{T(p_1)\cdots W_{h_j,w_j}(q_j) \cdots}_\Psi = \frac{h_j }{(z(p_1)-z(q_j))^2} \IP{\cdots W_{h_j,w_j}(q_j) \cdots}_\Psi+ \frac{\partial_{z(q_j)}\IP{W_{h_j,w_j}(q_j)\cdots}_\Psi}{z(p_1) - z(q_j)} +\regular\, .
\end{equation}
\begin{align} 
\nonumber     \IP{ \cW_3(p_1)\cdots W_{h_j,w_j}(q_j) \cdots}_\Psi = &\frac{w_j }{(z(p_1)-z(q_j))^3} \IP{ W_{h_j,w_j}(q_j) \cdots}_\Psi\\
\nonumber       &+ \frac{1}{(z(p)-z(q))^2} \IP{ W_{-1} W_{h_j,w_j}(q_j)\cdots}_\Psi\\
      &+ \frac{1}{z(p_1) - z(q_j)}\IP{ W_{-2} W_{h_j,w_j}(q_j)\cdots}_\Psi+\regular\, .
\end{align}

Including the Heisenberg algebra and its primaries $V_{\alpha_i}(s_i)$, the conformal blocks  $\Conf(C,W_{3,c} \otimes \Heis; W_{h_1,w_1} V_{\alpha_1}(g_1) \cdots W_{h_k,w_k} V_{\alpha_k}(g_k))$ can be defined similarly using \eqref{eq:w3-primary-ope-T}, \eqref{eq:w3-primary-ope-w3} and \eqref{eq:heisenberg-primary-ope}. 

\section{Branched free-field constraction}
In this section, we review the branched free-field construction which gives a recipe for correlation functions
\begin{equation} \label{eq:heisenberg-virasoro-correlator}
\IP{\cW_3(l_1)^{v_1}\cdots \cW_3(l_j)^{v_n}T(p_1)^{z_1}\cdots T(p_n)^{z_n} J(q_1)^{w_1} \cdots J(q_m)^{w_m} W_{\frac{1}{9},0}(b_1)\cdots W_{\frac{1}{9},0}(b_k)}_{\Psi} \, ,
\end{equation}
on $C$,
in terms of correlation functions
\begin{equation} \label{eq:heisenberg-correlator}
\IP{J(p_1)^{z_1}\cdots J(p_n)^{z_n}}_{\Psi}
\end{equation}
on $\tC$. Other primaries can also be inserted and we discuss insertions of the form $W_{h_{u,v},w_{u,v}}(p)$  and $W_{\beta^2,\frac{\sqrt{2}}{3}\beta^3}(p) V_{\beta }(p)$. The primaries $W_{\frac{1}{9},0}(b_1)\cdots W_{\frac{1}{9},0}(b_k)$ at branch points $b_i$ are usually known as twist fields and we will show how they appear naturally in the branched free-field construction.  

\subsection{Generators of VOA}\label{sec:generatorsmap}
The main ingredient in this recipe is a dictionary at the level of the local operators. 
Consider a point $p \in C$, which is not a branch point 
of $\pi$, and let $p^{(1)}$, $p^{(2)}$, $p^{(3)}$ be its three preimages
in $\tC$.
The dictionary is:
\begin{align} \label{eq:nab-operator-dictionary-J}
  &J(p) \rightsquigarrow 
  \frac{1}{\sqrt 3} \left( \tJ(p^{(1)}) + \tJ(p^{(2)}) + \tJ(p^{(3)}) \right) \, , \\
\label{eq:nab-operator-dictionary-T}  &T(p) \rightsquigarrow \frac16 (\nop{(\tJ(p^{(1)}) - \tJ(p^{(2)}))^2}+\nop{(\tJ(p^{(2)}) - \tJ(p^{(3)}))^2} +\nop{(\tJ(p^{(3)}) - \tJ(p^{(1)}))^2} ) \, 
\end{align}

\begingroup
\makeatletter
\renewcommand{\maketag@@@}[1]{\hbox{\m@th\normalsize\normalfont#1}}
\makeatother

{\scriptsize
\begin{align}
\label{eq:nab-operator-dictionary-W3}
\cW_3(p) \rightsquigarrow 
&-\frac1{9\sqrt{6}} \Bigg(\nop{(\tJ(p^{(1)}) +\tJ(p^{(2)})-2\tJ(p^{(3)}))(\tJ(p^{(2)})+\tJ(p^{(3)})-2\tJ(p^{(1)}) )(\tJ(p^{(1)})+\tJ(p^{(3)}) -2\tJ(p^{(2)}))}\Bigg)
\end{align}
}
\endgroup

We also define
\begin{equation}\label{Ttot}
  T^\tot(p)^z = T^\Heis(p)^z+T(p)^z\, ,
\end{equation}
and we get
\begin{equation}\label{Ttot}
  T^\tot(p)^z \rightsquigarrow \frac12 \nop{(\tJ(p^{(1)})^{z^{(1)}})^2 + (\tJ(p^{(2)})^{z^{(2)}})^2+(\tJ(p^{(3)})^{z^{(3)}})^2} \, .
\end{equation}

For convenience, we further define
\begin{equation}\label{W3tot}
  \cW_3^\tot(p)^z =\frac1{3\sqrt{2}} \nop{(J(p)^z)^3}+\cW(p)^z\, ,
\end{equation}
and we get
\begin{align} 
\label{eq:nab-operator-dictionary-W3tot}  \cW_3^\tot(p)^z \rightsquigarrow &\frac1{3\sqrt{6}} \Bigg(\nop{(\tJ(p^{(1)})^3 +\tJ(p^{(2)})^3+\tJ(p^{(3)})^3+6\tJ(p^{(1)})\tJ(p^{(2)})\tJ(p^{(3)}))}\Bigg) \, .
\end{align}
Note that $T^\tot$ and $\cW_3^\tot$ don't form a $W_3$-algebra.

This dictionary is to be understood as holding in correlation functions.
These formulas are invariant under 
the permutation $p^{(i)}\rightarrow p^{(j)}$.

\begin{center}
\begin{tikzpicture}
\usetikzlibrary{intersections}

\newcommand{\firstCol}{1.9}
\newcommand{\secondCol}{5.9}
\newcommand{\curveLowerY}{0.5}
\newcommand{\curveUpperY}{3.5}
\newcommand{\curveLowerZ}{2.0}
\newcommand{\curveUpperZ}{2.0}
\newcommand{\startX}{0.5}
\newcommand{\finalX}{7.5}
\newcommand{\curveHeight}{2.0}
\newcommand{\dashedHeight}{3.5}

\draw[ultra thick,\baseCurveColor] (\startX,0) -- (\finalX,0) node[right] {};

\draw[dashed,\vdashLineColor,name path=vertical1] (\firstCol,0) -- (\firstCol,\dashedHeight);
\draw[dashed,\vdashLineColor,name path=vertical2] (\secondCol,0) -- (\secondCol,\dashedHeight);

\draw[name path=curve1,thick,\coveringCurveColor] (\startX,\curveLowerY) .. controls (\firstCol,\curveHeight-1.5) and (\secondCol,\curveHeight+1.5) .. (\finalX,\curveUpperY);
\draw[name path=curve3,thick,\coveringCurveColor] (\startX,\curveUpperY) .. controls (\firstCol,\curveHeight+1.0) and (\secondCol,\curveHeight-1.0) .. (\finalX,\curveLowerY);
\draw[name path=curve2,thick,\coveringCurveColor] (\startX,\curveUpperZ) .. controls (\firstCol,\curveHeight+0.1) and (\secondCol,\curveHeight-0.1) .. (\finalX,\curveLowerZ);

\path [name intersections={of=curve1 and curve2, by=branch}];
\coordinate (branchb) at (branch |- 0,0);

\draw[dashed,\vdashLineColor] (branch |- 0,\dashedHeight) -- (branchb);

\node[dot,label=below:{\color{\tpColor}\( T(p_1) \)}] (Tp1) at (\firstCol,0) {};
\node[dot,label=below:{\color{\tpColor}\( T(p_2) \)}] (Tp2) at (\secondCol,0) {};

\path [name intersections={of=vertical1 and curve1, by=Jp11}];
\path [name intersections={of=vertical1 and curve2, by=Jp12}];
\path [name intersections={of=vertical1 and curve3, by=Jp13}];
\path [name intersections={of=vertical2 and curve1, by=Jp21}];
\path [name intersections={of=vertical2 and curve2, by=Jp22}];
\path [name intersections={of=vertical2 and curve3, by=Jp23}];

\node[dot,label={[xshift=1mm, yshift=2mm]left:{\color{\jpColor}\(  \tJ(p_1^{(1)}) \)}}] at (Jp11) {};
\node[dot,label={[xshift=1mm, yshift=3mm]left:{\color{\jpColor}\(  \tJ(p_1^{(2)}) \)}}] at (Jp12) {};
\node[dot,label={[xshift=1mm, yshift=7mm]left:{\color{\jpColor}\(  \tJ(p_1^{(3)}) \)}}] at (Jp13) {};
\node[dot,label={[xshift=-1mm, yshift=7mm]right:{\color{\jpColor}\( \tJ(p_2^{(1)}) \)}}] at (Jp21) {};
\node[dot,label={[xshift=-1mm, yshift=3mm]right:{\color{\jpColor}\( \tJ(p_2^{(2)}) \)}}] at (Jp22) {};
\node[dot,label={[xshift=-1mm, yshift=2mm]right:{\color{\jpColor}\( \tJ(p_2^{(3)}) \)}}] at (Jp23) {};

\drawbranchpointmarker{branchb};


\node at (\finalX+0.5,0) {\color{\baseCurveColor}\( C \)};
\node at (\finalX+0.5,\curveHeight) {\color{\coveringCurveColor}\( \tC \)};

\end{tikzpicture}
\end{center}

\subsection{Primary fields}\label{sec:primarymap}
We first consider the nonabelianization map for the $\cW_{3,c=2}$ primary $W_{h_{u,v},w_{u,v}}(p)$ on $C$, where $u,v \in \bbC$; see \eqref{huv} and \eqref{wuv}. In particular, when $u=1,v=0$, i.e. $\bm{\alpha}=\bm{\omega}_1$, $W_{\frac13,\pm\frac{\sqrt{2}}{9\sqrt{3}}}(p)$ is an important degenerate operator discussed in \cite{Kanno:2009ga,Coman:2017qgv,Dupic:2018pqr}. The conformal block with this degenerate field insertion satisfies 3rd order 
differential equations. This is analogous to the Virasoro case studied in \cite{Hao:2024vlg}, where the insertion of a $\Vir_{c=1}$ primary field $W_{\frac14}$ leads to a second order differential equation.

We require that $p$ is not a branch point of $\pi$. 
Then we introduce a dictionary extending \eqref{eq:nab-operator-dictionary-J},\eqref{eq:nab-operator-dictionary-T},\eqref{eq:nab-operator-dictionary-W3}: 
For a generic primary with conformal weight and $w$-charge defined in \eqref{huv} and \eqref{wuv}
\begin{equation} \label{eq:W-dictionary-gen}
 W_{h_{u,v},w_{u,v}}(p) \ \rightsquigarrow \ V_{-\frac{1}{3}(u+2v)}(p^{(i)})V_{-\frac13(u-v)}(p^{(j)}) V_{\frac13(2u+v)}(p^{(k)}),
\end{equation}
for arbitrary sheets $i,j,k$ which are all distinct.
This dictionary is engineered to produce the expected
analytic properties as the insertions $J(q)$, $T(q)$ and $W_3(q)$ approach
$W_{h_{u,v},w_{u,v}}(p)$. As an example we can check the OPE with $J(q)$ by 
using \eqref{eq:nab-operator-dictionary-J} and \eqref{eq:W-dictionary-gen}:
\begin{equation}
  J(q)  W_{h_{u,v},w_{u,v}}(p)\ \rightsquigarrow \  \frac{1}{\sqrt 3} \left( \tJ(p^{(1)}) + \tJ(p^{(2)}) + \tJ(p^{(3)}) \right) \, V_{-\frac{u+2v}{3}}(p^{(i)})V_{-\frac{(u-v)}3}(p^{(j)}) V_{\frac{(2u+v)}3}(p^{(k)})
\end{equation}
and further using 
\eqref{eq:heisenberg-primary-ope},
we check the right hand side is regular as $q \to p$,
matching the expectation that $J(q)  W_{h_{u,v},w_{u,v}}(p)$ is regular
as $q \to p$.
Similarly, we have
\begin{equation}
\begin{aligned}
  &T(q) W_{h_{u,v},w_{u,v}}(p)  \ \rightsquigarrow \\
  &\frac16 (\nop{(\tJ(p^{(1)}) - \tJ(p^{(2)}))^2}+\nop{(\tJ(p^{(2)}) - \tJ(p^{(3)}))^2} +\nop{(\tJ(p^{(3)}) - \tJ(p^{(1)}))^2} )\\
  &\times V_{-\frac{u+2v}{3}}(p^{(i)})V_{-\frac{(u-v)}3}(p^{(j)}) V_{\frac{(2u+v)}3}(p^{(k)})
\end{aligned}
\end{equation}
and in a local coordinate $z$ 
the right hand side has a singularity with leading term
$\frac{h_{u,v}}{(z(p)-z(q))^2}$. Similarly, we can check 
\begin{align}
  &\cW_{3}(q) W_{h_{u,v},w_{u,v}}(p)  \ \rightsquigarrow \\
  &\frac1{9\sqrt{6}} \Bigg(\nop{(\tJ(p^{(1)}) +\tJ(p^{(2)})-2\tJ(p^{(3)}))(2\tJ(p^{(1)}) -\tJ(p^{(2)})-\tJ(p^{(3)}))(\tJ(p^{(1)}) -2\tJ(p^{(2)})+\tJ(p^{(3)}))}\\
  &\times V_{-\frac{u+2v}{3}}(p^{(i)})V_{-\frac{u-v}3}(p^{(j)}) V_{\frac{2u+v}3}(p^{(k)})
\end{align}
has the expected singularity $\frac{w_{u,v}}{(z(p)-z(q))^3}$.

The dictionary
\eqref{eq:W-dictionary-gen} involves correlated 
insertions on all three sheets of $\tC$. For primaries of the form $W_{\beta^2,\frac{\sqrt{2}}{3}\beta^3}(p)$, we can also 
work with insertions on only one sheet of $\tC$ when we consider primaries on $C$ which are charged 
under both factors of $W_{3,c=1} \otimes \Heis$.
Namely, 
\begin{equation} \label{eq:chi-def}
\chi_\beta = W_{\beta^2,\frac{\sqrt{2}}{3}\beta^3}(p) V_{\beta }(p)
\end{equation}
on $C$.
By similar computations to the above, we can check that
this insertion can be obtained using the dictionary
\begin{equation} \label{eq:chi-dictionary}
\chi_\beta(p) \ \rightsquigarrow \ V_{\sqrt{3}\beta}(p^{(i)})
\end{equation}
for any choice of $i$. 
Note that when $\beta = \pm \frac1{\sqrt{3}}$, this corresponds to the degenerate operator mentioned in the beginning of \autoref{sec:primarymap} which are important to \autoref{sec:nonabelianizationmap}.
\subsection{Branch points}\label{sec:naivebp}
Compared to points $p$ that are not a branch point used in \autoref{sec:generatorsmap} and \autoref{sec:primarymap}, the branch points $b$ need a more careful study. We start by looking at $T^\tot(p)$. The right hand side of \eqref{Ttot} is a sum of three terms, each of which is finite because of the normal ordering.\footnote{If we studied $T$ instead of $T^\tot$, our life would be slightly harder at this moment, because there would be a cross-term which would have a singularity as $p \to b$; this is why we compute for $T^\tot$ instead.}
The local coordinates $z^{(1)}$, $z^{(2)}$ and $z^{(3)}$ all extend to good coordinate systems in $\tC$ 
including the point $\pi^{-1}(b)$. Finally,
there are no operators inserted at $\pi^{-1}(b)$.
We conclude that the right side is bounded 
as $p \to b$. Moreover, using $z^{(n)} = \E^{\frac{2\pi\ri}{3}n}z^{(1)}$, so $\de z^{(n)} / \de z^{(1)} = \E^{\frac{2\pi\ri}{3}n}$, we see that the zeroth order term in $z$ vanish automatically. Thus
$T^\tot(p)^z$ is bounded as $p \to b$ as $\regular\, z$.

In our case, $x = z^3$. Using
the rule \eqref{eq:virasoro-coordinate-change}
at $c=3$ and the relation $\{x,z\} = - \frac{4}{z^2}$ gives
\begin{equation}
  T^\tot(p)^{z} =(3z(p)^2)^2 T^\tot(p)^w + \frac{1}{4} \left(- \frac{4}{z(p)^2} \right) \, ,
\end{equation}
so finally
\begin{equation} \label{eq:T-branch-point-contributions-add}
  T^\tot(p)^x = \frac{1}{(3z(p)^2)^2} \left(  \frac{1}{ z(p)^2} + \regular\, z(p)\right) = \frac{1}{9 z(p)^6} + \frac{\regular}{ z(p)^3} = \frac{1}{9 x(p)^2} + \frac{\regular}{x(p)} \, ,
\end{equation}
which shows that $T^\tot(p)^x$ 
has a second-order pole at the branch point $b$,
with coefficient $\frac{1}{9}$.

To obtain the singularity of $T(p)^x$ from $T^\tot(p)^x$, we also consider the behavior of $J(p)$ near the branch point.
We switch to the common coordinate system $z^{(1)}$ for all three insertions
on $\tC$.
Then in this common coordinate
\begin{equation}
  J(p)^z \rightsquigarrow 
  \frac{1}{\sqrt 3} \left( \tJ(p^{(1)})^{z^{(1)}} +\E^{\frac{2\pi\ri}{3}} \tJ(p^{(2)})^{z^{(1)}}  +\E^{-\frac{2\pi\ri}{3}} \tJ(p^{(3)})^{z^{(1)}}  \right) \, ,
\end{equation}
with the relative $ \E^{\frac{2\pi\ri}{3}n}$ factor coming
from the fact that $z^{(n)} = \E^{\frac{2\pi\ri}{3}n}z^{(1)}$, so $\de z^{(n)} / \de z^{(1)} = \E^{\frac{2\pi\ri}{3}n}$.
It follows that $J(p)^z$ vanishes (to second order in $z$) as 
$z \to 0$. Changing coordinates to $x = z^3$ using
$J(p)^z = (3z(p)^2) J(p)^x$, we find that $J(p)^x$ is regular at $b$.

Finally, from the behavior of both $T^\tot$ and $J$ at the branch point
we can therefore deduce the behavior of $T = T^\tot - T^\Heis$: 
\begin{equation} \label{eq:T-singularity-branch-point}
  T(p)^x = \frac{1}{9 x(p)^2} + \frac{\regular}{x(p)} \, .
\end{equation}

The calculation for $W_3(p)^z$ is not as straight forward as for $T(p)^z$. Here we just show a check for the 0 $w$-charge by showing that $W_3(p)^x$ doesn't have a 3rd order singularity when $p\rightarrow 0$. To start, it is analogous to above to show that $W_3^\tot(p)^z$ is regular as $p \to b$ by checking additionally that 
\begin{equation}
\nop{6\tJ(p^{(1)})^{z(1)}\tJ(p^{(2)})^{z(2)}\tJ(p^{(3)})^{z(3)}}=\nop{6\tJ(z)^{z(1)}\tJ(z\E^{\frac{2\pi\ri}{3}})^{z(1)}\tJ(z\E^{-\frac{2\pi\ri}{3}})^{z(1)}}=\regular\ .
\end{equation}
As we discussed before $J(p)^x$ is regular at $b$. Thus $\cW_3(p)^z=\cW_3^{\tot}(p)^z-\frac1{3\sqrt{2}} \nop{(J(p)^z)^3}$ is also regular at $b$.
Under change of coordinate $x=z^3$,
\begin{align}\label{eq:W3-singularity-branch-point} 
\cW_3(p)^z=27 x^2\cW_3(p)^x.
\end{align}%
Thus $\cW_3(p)^x$ doesn't have a third order pole at $p=b$ from which we can conclude the $W_3$ conformal weight at the branch point $b$ is zero. The interpretation of \eqref{eq:T-singularity-branch-point}  and \eqref{eq:W3-singularity-branch-point} is that $\nab_0(\tPsi)$ is a conformal block 
with the primary field $W_{\frac{1}{9},0}$ inserted at each
branch point. This is consistent with the twist field insertion of conformal weight $\frac{1}{9}$ and $w$-charge 0 in \cite{Gavrylenko:2015cea}. In this work, we focus more on the conformal weight $\frac{1}{9}$ part.

In this work, we discuss a modification of the branched free-field construction described in this section, aiming at providing a construction which doesn't have the twist field insertion in the image.

\section{The nonabelianization map}\label{sec:nonabelianizationmap}

In this section, we generalize the nonablianization map in \cite{Hao:2024vlg} for the Virasoro$\otimes\Heis$ algebra at central charge $c=1$ to $W_3\otimes\Heis$ algebra at central charge $c=2$. We define a linear map
\begin{equation} \label{eq:nab-map}
 \nab_\cW: \quad \Conf(\tC,\Heis) \quad \to \quad \Conf(C,W_{3,c=2} \otimes \Heis) \, ,
\end{equation}
or the more general map with primary fields inserted
\begin{align} \label{eq:nab-map-with-primaries}
 \nab_\cW: \quad &\Conf(\tC,\Heis;V_{\sqrt{3}\beta_1}(p^{(i_1)}_1) \cdots V_{\sqrt{3}\beta_k}(p^{(i_k)}_k)) \quad \\
 &\to \quad \Conf(C,\cW_{3,c=2}\otimes \Heis ; 
 \chi_{\beta_1}(p_1) \cdots \chi_{\beta_2}(p_k)) \, .
\end{align}
Our construction uses the spectral network $\cW$ and the corresponding soliton data determined by $\cW$. And we claim that this construction results in the cancellation of the twist field insertions at branch points in the previous branched free-field construction descibed in \autoref{sec:naivebp}.

To be more specific, we give a definition 
of a $W_3$-Heisenberg block $\nab_\cW(\tPsi)$ on $C$,
beginning from a Heisenberg block $\tPsi$ on $\tC$,
using the extra operator $E_\ren(\cW)$ determined by the data of a spectral network $\cW$ on $C$ together with its corresponding soliton data and local coordinates
around branch points. To give some example,
the first few $W_3$ correlation functions in the conformal block $\nab_\cW(\tPsi)$ are
{\small \begin{align}
 \label{eq:nab-0-point} &\IP{1}_{\nab_\cW(\tPsi)} = \IP{E_\ren(\cW)}_{\tPsi},\\
\label{eq:nab-1-point}
\nonumber  &\IP{T(p)^z}_{\nab_\cW(\tPsi)}\\
  =& \frac{1}{6} \IP{ \nop{(\tJ(p^{(1)})^{z^{(1)}} - \tJ(p^{(2)})^{z^{(2)}})^2+(\tJ(p^{(2)})^{z^{(2)}} - \tJ(p^{(3)})^{z^{(3)}})^2+(\tJ(p^{(3)})^{z^{(3)}} - \tJ(p^{(1)})^{z^{(1)}})^2} \, E_\ren(\cW) }_\tPsi \, , \\
\label{eq:nab-2-point}
\nonumber  &\IP{T(p)^z T(q)^z}_{\nab_\cW(\tPsi)} \\
\nonumber  =& \frac{1}{36} \langle\nop{(\tJ(p^{(1)})^{z^{(1)}} - \tJ(p^{(2)})^{z^{(2)}})^2+(\tJ(p^{(2)})^{z^{(2)}} - \tJ(p^{(3)})^{z^{(3)}})^2+(\tJ(p^{(3)})^{z^{(3)}} - \tJ(p^{(1)})^{z^{(1)}})^2}\\
  &\nop{(\tJ(q^{(1)})^{z^{(1)}} - \tJ(q^{(2)})^{z^{(2)}})^2+(\tJ(q^{(2)})^{z^{(2)}} - \tJ(q^{(3)})^{z^{(3)}})^2+(\tJ(q^{(3)})^{z^{(3)}} - \tJ(q^{(1)})^{z^{(1)}})^2} \, E_\ren(\cW) \rangle_\tPsi \, , \\
\nonumber  &\IP{\cW_3(p)^z}_{\nab_\cW(\tPsi)}\\
\nonumber=&\frac1{9\sqrt{6}} \langle\nop{(\tJ(p^{(1)})^{z^{(1)}} +\tJ(p^{(2)})^{z^{(2)}} -2\tJ(p^{(3)})^{z^{(3)}})(2\tJ(p^{(1)})^{z^{(1)}}  -\tJ(p^{(2)})^{z^{(2)}}-\tJ(p^{(3)})^{z^{(3)}})\\
&(\tJ(p^{(1)})^{z^{(1)}}  -2\tJ(p^{(2)})^{z^{(2)}}+\tJ(p^{(3)})^{z^{(3)}})}\rangle_\tPsi\, .
\end{align}
}

\begin{figure}[htp]
     \centering
      \begin{subfigure}[b]{0.45\textwidth}
\begin{tikzpicture}
  \begin{scope}
    \coordinate (center) at (0,0);

    \draw [thick,\wallColor] (center) -- ++(120:1.5cm); 
    \draw [thick,\wallColor] (center) -- ++(240:1.5cm); 
    \draw [red,thick] (center) -- ++(0:1.5cm);   

    \drawbranchpointmarker{center};
   \draw [red,domain=18:342,line width=0.2mm] plot ({0.9*cos(\x)+0.4}, {0.4*sin(\x)});
    \draw[<-,red] (0.6,0.4 ) -- (0.61,0.4);
       \node[scale=0.6] at (0.6,0.4) [above] {$\color{red}j$};
        \node[scale=0.6] at (1.2,0.3) [above] {$\color{red}-\widetilde{\gamma_{ji}}$};
   \filldraw[black] (1.5,0) circle(1pt); 
   \node at (1.5,0) [right] {$z$};
   \node[scale=0.6] at (0.8,0) [above] {$\color{red}ij$};
   \draw[->] (-0.45,-0.45*1.732 ) -- (-0.451,-0.451*1.732);
      \draw[->] (-0.45,0.45*1.732 ) -- (-0.451,0.451*1.732);
         \node[scale=0.6] at (-0.45,0.45*1.732 ) [left] {$ji$};
         \node[scale=0.6] at (-0.45,-0.45*1.732 ) [left] {$ji$};
    \draw[->,red] (0.8,0 ) -- (0.9,0);
   \draw[snake it, orange] (-3, 0) to (0, 0);
  \draw[domain=0:0.5,smooth,variable=\t,red,line width=0.2mm]
    plot ({1.5-\t*\t},{\t*\t*\t});

  \draw[domain=0:0.5,smooth,variable=\t,red,line width=0.2mm]
    plot ({1.5-\t*\t},{-\t*\t*\t});
  \end{scope}
\end{tikzpicture}
\caption{}
\label{soln1}
\end{subfigure}
\hfill
      \begin{subfigure}[b]{0.45\textwidth}
\begin{tikzpicture}
  \begin{scope}
    \coordinate (center) at (0,0);

    \draw [thick,\wallColor] (center) -- ++(120:1.5cm); 
    \draw [thick,\wallColor] (center) -- ++(240:1.5cm); 
    \draw [red,thick] (center) -- ++(0:1.5cm);   

    \drawbranchpointmarker{center};
   \draw [red,domain=13:347,line width=0.2mm] plot ({0.5*cos(\x)+0.8}, {0.4*sin(\x)});
    \draw [red,domain=0:360,line width=0.2mm] plot ({0.3*cos(\x)}, {0.2*sin(\x)});
    \draw[<-,red] (0.9,0.4 ) -- (0.91,0.4);
      \draw[<-,red] (0,-0.2 ) -- (0.1,-0.2);
       \node[scale=0.6] at (0.9,0.4 ) [above] {$\color{red}j$};
        \node[scale=0.6] at (0.1,-0.2 ) [below] {$\color{red}j$};
        \node[scale=0.6] at (1.3,0.3) [above] {$\color{red}-\widetilde{\gamma_{ji}}'$};
   \filldraw[black] (1.5,0) circle(1pt); 
   \node at (1.5,0) [right] {$z$};
   \node[scale=0.6] at (0.8,0) [above] {$\color{red}ij$};
   \draw[->] (-0.45,-0.45*1.732 ) -- (-0.451,-0.451*1.732);
      \draw[->] (-0.45,0.45*1.732 ) -- (-0.451,0.451*1.732);
         \node[scale=0.6] at (-0.45,0.45*1.732 ) [left] {$ji$};
         \node[scale=0.6] at (-0.45,-0.45*1.732 ) [left] {$ji$};
    \draw[->,red] (0.8,0 ) -- (0.9,0);
   \draw[snake it, orange] (-3, 0) to (0, 0);
   
  \draw[domain=0:0.46,smooth,variable=\t,red,line width=0.2mm]
    plot ({1.5-\t*\t},{\t*\t*\t});

  \draw[domain=0:0.46,smooth,variable=\t,red,line width=0.2mm]
    plot ({1.5-\t*\t},{-\t*\t*\t});

  \end{scope}
\end{tikzpicture}
\caption{}
\label{soln2}
\end{subfigure}
\caption{Two possible detour paths for the regularization given by soliton charges $-\widetilde{\gamma_{ji}}\in\widetilde{C}$. 
 }
\label{2soln}
\end{figure}
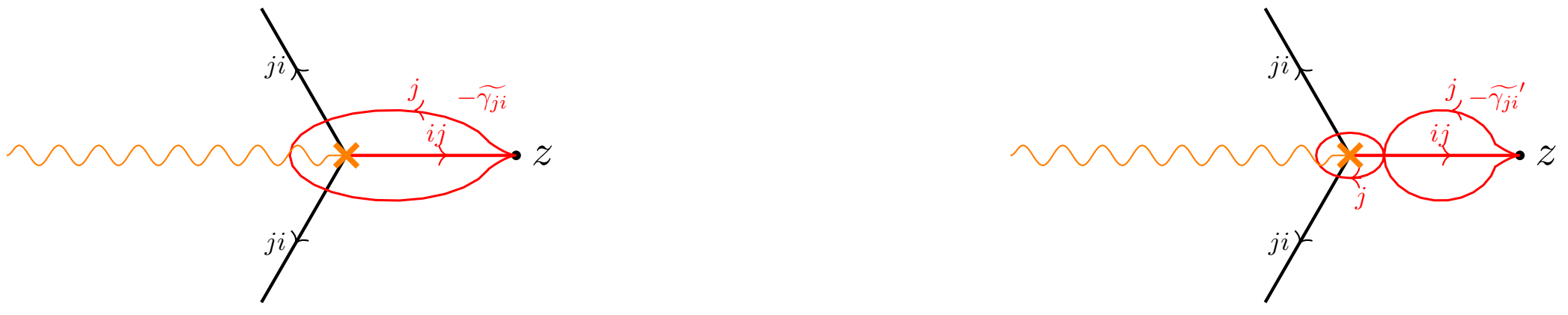

Other $n$-point functions are defined similarly,
using the dictionary \eqref{eq:nab-operator-dictionary-T} and \eqref{eq:nab-operator-dictionary-W3} for each operator inserted on $C$ together with inserting the extra operator $E_\ren(\cW)$ in each correlator. The operator $E_\ren(\cW)$ is 
given as follows. First define 
 \begin{equation} \label{eq:EWreg-def}
E(\cW) =\exp \left[ -\frac{1}{2 \pi \I} \int_\cW\mu(-\widetilde{\gamma_{ji}}^{cb}) 
\tikz[remember picture, baseline]{
  \node[anchor=base, inner sep=0pt] (psiplus) {$\psi_+(q^{(+)})^{z^{(+)}}$};
}
\, 
\tikz[remember picture, baseline]{
  \node[anchor=base, inner sep=0pt] (psiminus) {$\psi_-(q^{(-)})^{z^{(-)}}$};
}
\, \de z \right]  \, ,
\end{equation}
\begin{tikzpicture}[overlay, remember picture]
  \coordinate (start) at ($(psiplus.south)!0.5!(psiplus.north)$);
  \coordinate (end) at ($(psiminus.south)!0.5!(psiminus.north)$);
  \coordinate (startShifted) at ([yshift=-2ex]start);
  \coordinate (endShifted) at ([yshift=-2ex]end);
  \draw[thick] (startShifted) -- (endShifted);
  \draw[thick] (startShifted) -- ++(0,0.5ex);
  \draw[thick] (endShifted) -- ++(0,0.5ex);
  \coordinate (mid) at ($(startShifted)!0.5!(endShifted)$);
  \node at ($(mid)+(0,-1.5ex)$) {$\ell_\cG(q)$};
\end{tikzpicture}
where $-\widetilde{\gamma_{ji}}^{cb}$ is the lift of a soliton charge $-\widetilde{\gamma_{ji}}$, for example as shown in \autoref{2soln}, to the circle bundle $\widetilde{C}^{cb}$ of tangent directions to $\widetilde{C}$ and $\mu(-\widetilde{\gamma_{ji}}^{cb})$ is the corresponding soliton degeneracy (which is closely related to Stokes multiplier).  \autoref{2soln} is drawn near a branch point $b$ of type $(ij)$, where the two choices of a soliton path are shown corresponding to different orientation. However, for a (123) type branch point that we study in this paper, there is a more natural choice using the smallest winding angle for the orientation. We will keep this as our convention. $\ell_\cG(q)$ represents the leash introduced in \cite{Hao:2024vlg} and we make the choice using again the natural winding direction. Note that at the first glance, our definition of $E(\cW)$ is not a straight forward generalization of the ones in the Virasoro case \cite{Hao:2024vlg} since \eqref{eq:EWreg-def} depends on the soliton data, which doesn't appear explicitly in \cite{Hao:2024vlg}. In the later part, we will discuss that \eqref{eq:EWreg-def} is the correct generalization that we need.
The renomalized $ E_\ren(\cW) $ of \eqref{eq:EWreg-def} is needed since correlation functions 
involving $ E(\cW)$ diverge when $\eps \to 0$. We propose
 \begin{equation} \label{eq:renormalized-correlator-general}
 E_\ren(\cW) = \lim_{\eps \to 0} \eps^{\frac{k_1}{8}} \eps^{\frac{k_2}{3}}  E(\cW)  \, ,
\end{equation}
where $k_1$ is the number of branch points of (ij) type and $k_2$ is the number of branch points of (123) type. The conjecture for the (ij) type branch points is due to the similarity to the Virasoro case. However in this work, we won't discuss this further.
In this paper we focus on the case where the branch points are all of (123) type, thus we use the normalization
\begin{equation} \label{eq:renormalized-correlator}
 E_\ren(\cW) = \lim_{\eps \to 0} \eps^{\frac{k}{3}}  E(\cW)  \, ,
\end{equation}
where $k$ is the number of branch points of (123) type. This dependence on $\eps^{-\frac{k}{3}}$ of correlation functions is a local behavior and we can understand it from 
what happens in the neighborhood of a branch point. In fact this local behavior can be proven 
by studying a simple model example. We do this in \autoref{sec:tau} below.

Correlation functions involving $E_\ren(\cW)$
are well defined and (generically) nonzero.
 However, $E_\ren(\cW)$ depends on the choice of 
local coordinate $y$ around each branch point, with scaling dimension $-\frac{1}{9}$, i.e.
\begin{equation}
  E^y_\ren(\cW) = \left\lvert \frac{\de y'(b)}{\de y(b)} \right\rvert^{- \frac{1}{9}} E^{y'}_\ren(\cW)  \, .
\end{equation}
In fact, such an insertion at each branch point with 
dimension $-\frac{1}{9}$ is
what is needed to cancel the known weight $\frac{1}{9}$ of twist field in the branched free-field construction.

With these definitions, we built a nonabelianization map \eqref{eq:nab-map} for conformal blocks such that the correlation functions $\IP{\cdots}_{\nab_\cW(\tPsi)}$ have all the expected properties for a conformal block on $C$.

In the above discussion, we have given a recipe of the generalization of the nonabelianization map from $\Vir_{c=1}$ to $W_{3,c=2}$. In the following, we give a motivation on why we expect the nonabelianization map defined in \cite{Hao:2024vlg} for the Virasoro algebra at central charge $c=1$ can be generalized to the $W_3$ algebra at central charge $c=2$.

An important calculation in the nonabelianization map of $\Vir_{c=1}$ is for the conformal block $ \langle 1 \rangle_\Psi\in \Conf(C,\Vir_{c=1} \otimes \Heis)$ in the simplest example  \cite{Hao:2024vlg}. This correlator belongs to a general type of correlators which have the meaning of the isomonodromic $\tau$ functions in some integrable system when they diagonalize Verlinde loop operators. These $\tau$ functions capture the isomonodromic deformations of second order differential equations which can be obtained by taking the Nekrasov-Shatashvili (NS) limit of the BPZ equations. The AGT correspondence further relates those differential equations  to 4d $N=2$ theories as the quantization of Seiberg-Witten curves. 
A basis of solutions to such differential equations are given by Nekrasov partition functions with canonical surface defects -- which by AGT correspondence are identified with conformal blocks with a degenerate field $W_{\frac14}$ insertion, both in the Nekrasov-Shatashvili limit. 

In the nonabelianization map of conformal blocks developed in \cite{Hao:2024vlg}, a crucial role is played by spectral networks -- also known as Stokes graphs. In fact, the explicit expression of the nonablianization map utilizing spectral networks is partly motivated by their ability to capture the Stokes data of second order differential equations. In $\Vir_{c=1}\otimes \Heis$, \cite{Hao:2024vlg} studies a double cover $\widetilde{C}$ of $C$. Similar to \eqref{eq:chi-def} and \eqref{eq:chi-dictionary}, the  degenerate field $\chi_{\pm\frac12}=W_{\frac14}V_{\pm\frac1{\sqrt{2}}}$ is mapped to 
\begin{equation} \label{eq:chi-dictionary-vir}
  \chi_{\pm\frac12}(p) \ \rightsquigarrow \ V_{\pm1}(p^{(i)})
\end{equation}
on the double cover $\widetilde{C}$, where $i$ is either sheet of $\widetilde{C}$.

Analogously, in this paper we will calculate a key example of the conformal block $ \langle 1 \rangle_\Psi\in \Conf(C,\cW_{3,c=2} \otimes \Heis)$ in \autoref{sec:tau}. We conjecture that in the $W_3$ case considered in this paper, we should consider the third order differential equation in the Nekrasov-Shatashvili limit of the $W_3$-BPZ equations \cite{Dupic:2018pqr,Lafay:2024tne} satisfied by the simplest
nontrivial degenerate field, $W_{\frac13,\pm\frac{\sqrt{2}}{9\sqrt{3}}}(p)$ of $W_{3,c=2}$ and the corresponding $W_{3,c=2} \otimes \Heis$ primary $W_{\frac13,\pm\frac{\sqrt{2}}{9\sqrt{3}}}V_{\pm\frac1{\sqrt{3}}}$. By \eqref{eq:chi-def} and \eqref{eq:chi-dictionary}, the insertion needed is also $V_{\pm 1}(p^{(i)})$, where $i$ denotes any sheet of $\tC$ the projection map $\pi: \tC  \xrightarrow{3:1} C$. 
Since the insertion on the cover corresponding to the differential equation is also the free fermions $V_{\pm 1}(p^{(i)})$, we expect a straight forward generalization of the nonabelianization map of the Virasoro $c=1$ case \cite{Hao:2024vlg}. 

To further argue why we need the spectral network and soliton data in the formula of \eqref{eq:EWreg-def}, in this section we will provide supporting evidences in terms of Stokes phenomenon. For later convenience, we first review some basics of the Virasoro work in \cite{Hao:2024vlg} where the operator $E(\cW)$ is defined as
\begin{equation} \label{eq:bilocal-wall-operator}
E(\cW) =\exp \left[ -\frac{1}{2 \pi \I} \int_\cW 
\tikz[remember picture, baseline]{
  \node[anchor=base, inner sep=0pt] (psiplus) {$\psi_+(q^{(+)})^{z^{(+)}}$};
}
\, 
\tikz[remember picture, baseline]{
  \node[anchor=base, inner sep=0pt] (psiminus) {$\psi_-(q^{(-)})^{z^{(-)}}$};
}
\, \de z \right]  \, .
\end{equation}
\begin{tikzpicture}[overlay, remember picture]
  \coordinate (start) at ($(psiplus.south)!0.5!(psiplus.north)$);
  \coordinate (end) at ($(psiminus.south)!0.5!(psiminus.north)$);
  \coordinate (startShifted) at ([yshift=-2ex]start);
  \coordinate (endShifted) at ([yshift=-2ex]end);
  \draw[thick] (startShifted) -- (endShifted);
  \draw[thick] (startShifted) -- ++(0,0.5ex);
  \draw[thick] (endShifted) -- ++(0,0.5ex);
  \coordinate (mid) at ($(startShifted)!0.5!(endShifted)$);
  \node at ($(mid)+(0,-1.5ex)$) {$\ell_\cG(q)$};
\end{tikzpicture} 

In the following, we give an argument for the inclusion of soliton degeneracies in our $W_3$-algebra case. They are closely related to the discontinuities that we expect for conformal blocks with degenerate fields insertions. Without already diving into the concrete computation, we sketch an argument, where we ignore the leash. However, one can also do the concrete calculation similarly.
Regarding the organization, we first show that the inclusion of soliton degeneracies is consistent with the description in \cite{Hao:2024vlg}; see \autoref{sec:disc_vir}. Then in \autoref{sec:disc_W3}, we will calculate the soliton degeneracies for $\cW_{3,c=2}$ conformal blocks which we need in the construction of the nonabelianization map. 

 \subsection{\texorpdfstring{$c=1$ Virasoro blocks with degenerate field and discontinuities}
{c=1 Virasoro blocks with degenerate field and discontinuities}}\label{sec:disc_vir}
In \cite{Hao:2024vlg}, it is observed that the 
the normalized fermion 2-point function on $\tC$ with the spectral network inserted
\begin{equation}\label{eq:deffzw}
F(z,w) = \frac{ \langle
\tikz[remember picture, baseline]{
  \node[anchor=base, inner sep=0pt] (psiplus1) {$\psi_+(v)$};
}
\,
\tikz[remember picture, baseline]{
  \node[anchor=base, inner sep=0pt] (psiminus1) {$\psi_-(w)$};
}
\, E_\ren(\cW) \rangle_\tPsi }{ \langle E_\ren(\cW) \rangle_\tPsi} \, ,
\end{equation}
 which is the preimage of conformal blocks with the degenerate fields \eqref{eq:chi-dictionary-vir} insertion,
reflects a very similar Stokes phenomenon of a second order differential equation which can be obtained as the NS limit of the corresponding BPZ equation\footnote{For the simple example that we talk about below, there is actually no corresponding differential equation, but we can study the soliton degeneracies corresponding to the specific spectral network as normal.}.
We explain the discontinuities of conformal blocks in the following, and discuss the corresponding Stokes phenomena in \autoref{mathmon}. In fact, \autoref{mathmon} is usually used in the exact WKB analysis of differential equations. There is an alternative description in the 4d $N=2$ gauges theories which uses the language of solitons supported by canonical surface defects and the discontinuity can be expressed in terms of the soliton degeneracy. We show the 4d $N=2$ description and how it matches with the one in former one in \autoref{sec:virsol}.

In \cite{Hao:2024vlg}, it is shown that the normalized two point function $F(z,w) $ defined by \eqref{eq:deffzw} is piecewise analytic: it jumps by addition of $- \I F(-v,w)$ when $v$ crosses $\cW^{(-)}$ in the counterclockwise direction,
and jumps by addition of $-\I F(v,-w)$ when $w$ crosses $\cW^{(+)}$ in the counterclockwise direction. For concreteness, we focus on the example of the Riemann surface $C = \bbC\bbP^1$, and the double cover $\tC = \bbC\bbP^1$, with the double cover map
\begin{align}\label{toyVir}
\pi: \tC & \to C \\
z & \mapsto x = z^2.
\end{align}
The exponent of \eqref{eq:bilocal-wall-operator} under the change of coordinates can be written as
\begin{equation} \label{eq:EW-rewritten}
 E(\cW) = \exp \left( \frac{1}{2 \pi} \int_{\cW^{(+)}} \de z \,
 \tikz[remember picture, baseline]{
   \node[anchor=base, inner sep=0pt] (psiplus) {$\psi_+(z)$};
 }
 \,
 \tikz[remember picture, baseline]{
   \node[anchor=base, inner sep=0pt] (psiminus) {$\psi_-(-z)$};
 }
 \right) \, .
\end{equation}
The discontinuity can be further checked by using a crucial relation in \cite{Hao:2024vlg}, that is the normalized $2n$-fermion correlation functions are determinants of matrices of normalized $2$-fermion correlation functions:
\begin{samepage}
\begin{equation} \label{eq:free-fermion-determinant-formula}
  \frac{  \left\langle  \prod_{i=1}^n 
    \tikz[remember picture, baseline]{
        \node[anchor=base, inner sep=0pt] (psiplus_num) {$\psi_+(z_i)$};
    }
    \,
    \tikz[remember picture, baseline]{
        \node[anchor=base, inner sep=0pt] (psiminus_num) {$\psi_-(w_i)$};
    }
    \right\rangle_{\tPsi} }{\IP{1}_\tPsi}= \det \left[ 
    \frac{
    \IP{
    \tikz[remember picture, baseline]{
        \node[anchor=base, inner sep=0pt] (psiplus_rhs) {$\psiplus(z_i)$};
    }
    \, 
    \tikz[remember picture, baseline]{
        \node[anchor=base, inner sep=0pt] (psiminus_rhs) {$\psi_-(w_j)$};
    }
    }_\tPsi}{\IP{1}_\tPsi}
    \right]_{i,j=1}^{n} \, .
\end{equation}
\begin{tikzpicture}[overlay, remember picture]
    \coordinate (start1) at (psiplus_num.north);
    \coordinate (end1) at (psiminus_num.north);
    \coordinate (start1Shifted) at ([yshift=1ex]start1);
    \coordinate (end1Shifted) at ([yshift=1ex]end1);
    \draw[thick] (start1Shifted) -- (end1Shifted);
    \draw[thick] (start1Shifted) -- ++(0,-0.5ex);
    \draw[thick] (end1Shifted) -- ++(0,-0.5ex);

    \coordinate (start2) at (psiplus_rhs.north);
    \coordinate (end2) at (psiminus_rhs.north);
    \coordinate (start2Shifted) at ([yshift=1ex]start2);
    \coordinate (end2Shifted) at ([yshift=1ex]end2);
    \draw[thick] (start2Shifted) -- (end2Shifted);
    \draw[thick] (start2Shifted) -- ++(0,-0.5ex);
    \draw[thick] (end2Shifted) -- ++(0,-0.5ex);
\end{tikzpicture}\noindent\unskip 
\end{samepage}
We show $v\rightarrow{\mathcal{W}^{(+)}}$ as an example:
\begin{align}
\nonumber&\quad\disc_{w\rightarrow \cW^{(+)}}F(v,w)\\
\nonumber&=\disc_{w\rightarrow \cW^{(+)}}\left(\psi_+(v)\psi_-(w)\exp \left( \frac{1}{2 \pi} \int_{\cW^{(+)}}\psi_+(z)\psi_-(-z) \de z \right)\right)\\
\nonumber&=\disc_{w\rightarrow \cW^{(+)}}\left( \sum_{n=0}^\infty \left(\frac{-1}{2\pi}\right)^n \int_{\cW^{(+)}}\cdots\int_{\cW^{(+)}} \frac{1}{v+x_1}\left(\prod_{i=1}^{n-1}\frac{1}{x_i+x_{i+1}}\right)\frac{1}{x_n-w} \, \de x_1\cdots\de x_n\right)\\
\nonumber&=\disc_{w\rightarrow \cW^{(+)}}\left( -\ri\sum_{n=1}^\infty \left(\frac{-1}{2\pi}\right)^{n-1} \int_{\cW^{(+)}}\cdots\int_{\cW^{(+)}} \frac{1}{-v+x_2}\left(\prod_{i=2}^{n-1}\frac{1}{x_i+x_{i+1}}\right)\frac{1}{x_{n}-w} \, \de x_2\cdots\de x_{n}\right)\\
&=-\ri\psi_+(-v)\psi_-(w), 
\end{align}
where the discontinuity is defined as $v$ in the counterclockwise minus the clockwise direction.
Note that here $w,v$ are points on $\tC $. Let $x=v^2$ be the point on $C$. If we view $\psi_+(v) $ as $\psi^{(1)}(x) $ and  $\psi_+(-v) $ as $\psi^{(2)}(x) $, this discontinuity is the familiar form of the Voros connection formula.

We will first discuss that the Virasoro example in \cite{Hao:2024vlg} also fits in the proposed general formula \eqref{eq:EW-secsol}, with a proper translation of language.

\subsubsection{Exact WKB description from math}\label{mathmon}
The Virasoro conformal blocks correspond to the 4d $N=2$ supersymmetric theories  by the AGT correspondence. And those theories can be described by second order ordinary differential equations in the NS phase
\begin{equation}
\hbar^2\partial_x^2\psi(x)-P(x)\psi(x)=0.
\end{equation}
$\hbar$ here is related to a parametrization $b=\sqrt{\frac{\epsilon_1}{\epsilon_2}}$ in the limit $\epsilon_2\rightarrow 0$ while $\epsilon_1=\hbar$ is fixed. In this setup, the spectral networks are identified with the Stokes graphs of the second order differential equations. Start with solutions $(\psi^{(1)},\psi^{(2)})$ obtained by the WKB ansatz 
\begin{equation}\label{borelsolwkb}
\psi^{\text{WKB},(i)}(x)=\exp\left(\frac{1}{\hbar}\text{Reg}\left(\int_{b}^xY^{(i)}(x,\hbar) \de x\right)\right),
\end{equation}
where
\begin{equation}\label{wkbY}
Y^{(i)}(x,\hbar) = \sum_{n=0}^\infty Y^{(i),n}(x)\hbar^{n}\, 
\end{equation}
is defined for $i=1,2$ determined by a choice of $ Y^{(i),0}=\pm\sqrt{P(x)}$ using the principal branch of the square root. In this work, we define $ Y^{(1),0}=\sqrt{P(x)}$.
 We can look at the even and odd part
\begin{align} 
\label{eodef} Y_{\rm even}=& \sum_{n=0}^\infty Y^{2n}(x)\hbar^{2n}\, ,\\
Y_{\rm odd}=&\sum_{n=0}^\infty Y^{2n+1}(x)\hbar^{2n+1},
\end{align}
separately. They further define even and odd part $\psi^{\text{even}}$ and $\psi^{\text{odd}}$ of the solution in a similar way as \eqref{borelsolwkb}. $Y_{\rm odd}$ and $Y_{\rm even}$ are related by
\begin{equation} \label{oddeven}Y_{\rm odd} = -{\frac{\hbar}{2}}{\frac{\rd}{\rd x}}\log Y_{\rm even}.
\end{equation}
So it is convenient to define
\begin{equation}\label{wkbsolreg}
\psi^{\text{WKB},(i)}(x)=\frac{1}{\sqrt{s(Y^{(1)}_\even(\epsilon)})}\exp\left(\frac{1}{\hbar}\text{Reg}\left(\int_{b}^xY^{(i)}_\even(x,\hbar) \de x\right)\right)(\epsilon),
\end{equation}
where Reg in \eqref{wkbY} stands for a regularization of the normalization at a branch point which is needed for even terms. 
In the exact WKB, a commonly used regularization is given by using a soliton charge as integral contour and then take $\frac12$ of the integrals. There are two possible choices as shown in \autoref{2soln}. The difference affects 
\begin{equation}
\label{1looppart}\E^{\frac{1}{\hbar}\oint_{\gamma_c} Y_{\rm odd}}= \E^{-\frac{1}{\hbar}\frac{\hbar}{2} \oint \de \log\sqrt{P} },
\end{equation}
where the first path gives $= \E^{-\frac{\pi}{2} \ri}=-\ri$ and the second one gives $= \E^{-\frac{\pi}{2} \ri}=\ri$.
But the definition \eqref{wkbsolreg} is defined by even part, the two choices lead to the same result.
The series $\text{Reg}(Y)$ is still an asymptotic series, a well-know resummation is given by the borel summation denoted by $s(\text{Reg}(Y))$. We thus define an actual solution by
\begin{equation}\label{borelsol}
\psi^{(i)}(x)=\frac{1}{\sqrt{s(Y^{(1)}_\even(\epsilon)})}\exp\left(\frac{1}{\hbar}s\left(\text{Reg}\left(\int_{b}^xY^{(i)}_\even(x,\hbar) \de x\right)\right)(\epsilon)\right).
\end{equation}

Let's take a closer look at this basis of solution $(\psi^{(1)},\psi^{(2)})$ defined in \eqref{borelsol}. At a wall of type $ij$, the even part of the solutions $(\psi^{(1)}_{cc},\psi^{(2)}_{cc})$ to the counterclockwise direction of the wall can be expressed in terms of the ones in the clockwise direction $(\psi^{(1)}_{c},\psi^{(2)}_{c})$ as 

\begin{align}\label{jumpuni}
\psi^{(i)}_{cc}&=\psi^{(i)}_{c}-\ri\psi^{(j)}_{c},\\
\psi^{(j)}_c&=\psi^{(j)}_{cc};
\end{align}
see \autoref{fig:Vorosconn}. This discontinuity across the walls of Stokes graphs or spectral networks is also known as a wall crossing formula. Around a loop $\gamma_C\in C$ as shown in \autoref{fig:Vorosconn} which follows the counterclockwise direction surrounding the branch point once, we can compute the monodromy matrix of solutions $\begin{pmatrix}\psi^{(1)}, & \psi^{(2)}\end{pmatrix}$ around a branch point as 
\begin{align}
M_{\gamma_c}=\begin{pmatrix}
1 & 0 \\
-\ri & 1
\end{pmatrix}.
\begin{pmatrix}
1 & -\ri \\
0 & 1
\end{pmatrix}.
\begin{pmatrix}
1 & 0 \\
-\ri & 1
\end{pmatrix}.
\begin{pmatrix}
0 & 1 \\
1 & 0
\end{pmatrix}
=
\left(
\begin{array}{cc}
 -\ri & 0 \\
 0 & -\ri \\
\end{array}
\right)
\end{align}
or
\begin{align}
M_{\gamma_c}=\begin{pmatrix}
1 & -\ri \\
0 & 1
\end{pmatrix}.
\begin{pmatrix}
1 & 0 \\
-\ri & 1
\end{pmatrix}.
\begin{pmatrix}
1 & -\ri \\
0 & 1
\end{pmatrix}.
\begin{pmatrix}
0 & 1 \\
1 & 0
\end{pmatrix}=\left(
\begin{array}{cc}
 -\ri & 0 \\
 0 & -\ri \\
\end{array}
\right)
\end{align}
Thus
\begin{equation}
\begin{pmatrix}\psi^{(1),\text{even}}, & \psi^{(2),\text{even}}\end{pmatrix}.M_{\gamma_c}=\begin{pmatrix}-\ri\psi^{(1)}, & -\ri\psi^{(2)}\end{pmatrix}
\end{equation}

Note that this monodromy matches with the expected one coming from the factor $\frac{1}{\sqrt{s(Y^{(1)}_\even(\epsilon)})}$ 
In particular, it just comes from the leading
term in $Y_{\mathrm{even}}$ when approaching the branch point, i.e. it is the same as the $\frac{1}{\sqrt{\E^{2\pi\ri}}}=-\ri$ since $\sqrt{P}$ is single-valued on $\Sigma$.

\subsubsection{Alternative description from physics}\label{sec:virsol}
We can also study the Stokes phenomenon and wall crossing formulas for the ordinary differential equations from their corresponding 4d $N=2$ theories \cite{Gaiotto:2012rg}. 
In 4d $N=2$ theories, the object studied is $\Psi^{(i)}(z)$ which should be thought of as
\begin{equation}
\Psi^{(i)}(x)=\exp\left(\frac{1}{\hbar}s\left(\text{Reg}\left(\int_{b}^xY^{(i)}(x,\hbar) \de x\right)\right)(\epsilon)\right).
\end{equation}
The discontinuity at a wall ij is described by
\begin{equation}\label{sec2:jumpsol2}
\begin{aligned} 
&{\rm disc}(\Psi^{(i)})(z)=\mu(-\widetilde{\gamma_{ji}}^{cb})\Psi^{(j)}(z)
\end{aligned}
 \end{equation}
where $\mu(-\widetilde{\gamma_{ji}}^{cb})$ is the soliton degeneracy. 
Note that in this case, the lifts of the two $-\widetilde{\gamma_{ji}}^{cb}$ in \autoref{2soln}  are different by $H$ where $H\in H_1(\widetilde{C}^{cb};\mathbb{Z})$ is a path winds once around a fiber of $\widetilde{C}^{cb}$ counterclockwise. 
The difference is a minus sign for $\pm \ri$.
As explained in \cite{Gaiotto:2012rg}, the soliton degeneracies satisfy
\begin{equation}
\mu(\widetilde{\gamma_{ji}}^{cb})/\mu(\widetilde{\gamma_{ji}}^{cb'})=(-1)^{w(\widetilde{\gamma_{ji}}^{cb},\widetilde{\gamma_{ji}}^{cb'})}
\end{equation}
where $w$ is the winding number. Combing this difference of soliton degeneracies by a minus sign with the minus sign coming from the integral of the odd terms for different soliton paths $-\widetilde{\gamma_{ji}}$ lead to the same result as \eqref{jumpuni}.

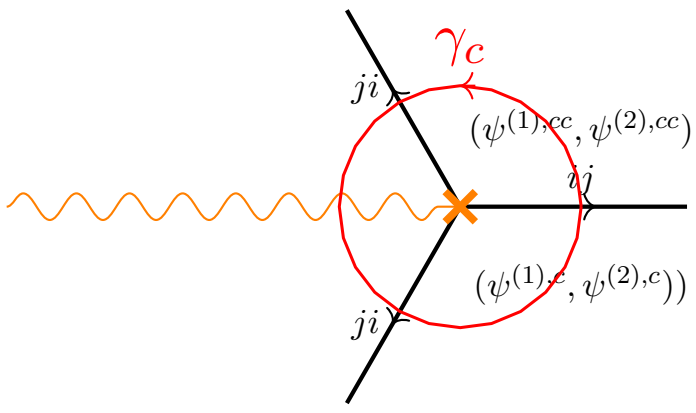
\begin{figure}
\centering
\begin{tikzpicture}
  \begin{scope}
    \coordinate (center) at (0,0);

    \draw [thick,\wallColor] (center) -- ++(120:1.5cm); 
    \draw [thick,\wallColor] (center) -- ++(240:1.5cm); 
    \draw [thick] (center) -- ++(0:1.5cm);   

    \drawbranchpointmarker{center};
   \node[scale=0.6] at (0.8,0) [above] {$ij$};
   \draw[->] (-0.45,-0.45*1.732 ) -- (-0.451,-0.451*1.732);
      \draw[->] (-0.45,0.45*1.732 ) -- (-0.451,0.451*1.732);
         \node[scale=0.6] at (-0.45,0.45*1.732 ) [left] {$ji$};
         \node[scale=0.6] at (-0.45,-0.45*1.732 ) [left] {$ji$};
    \draw[->] (0.8,0 ) -- (0.9,0);
   \draw[snake it, orange] (-3, 0) to (0, 0);
\node[scale=0.6] at (0.8,0.3) [above] {$(\psi^{(1),cc},\psi^{(2),cc})$};
\node[scale=0.6] at (0.8,-0.3) [below] {$(\psi^{(1),c},\psi^{(2),c}))$};
    \draw [red,domain=0:360,line width=0.2mm] plot ({0.8*cos(\x)}, {0.8*sin(\x)});
    \node at (0,0.8) [above,red] {$\gamma_c$};
        \draw[<-,red] (0,0.8 ) -- (0.1,0.8);
  \end{scope}
\end{tikzpicture}
\caption{Local spectral network for a branch point of differential equations corresponding to the 4d $N=2$ theories.}
\label{fig:solsconn}
\end{figure}

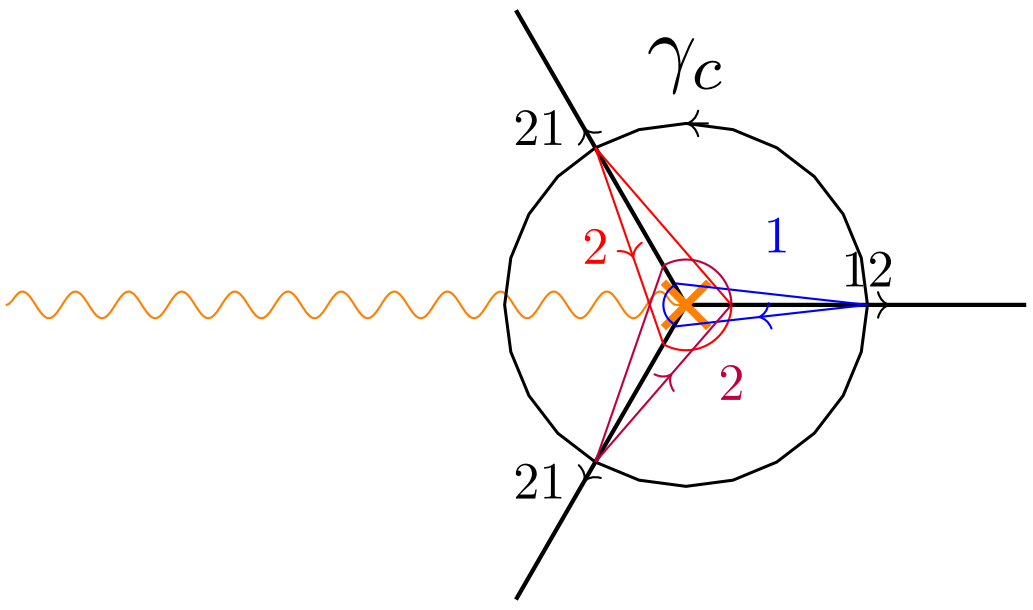
\begin{figure}
\centering
\begin{tikzpicture}[scale=1.5, transform shape]
  \begin{scope}
    \coordinate (center) at (0,0);

    \draw [thick,\wallColor] (center) -- ++(120:1.5cm); 
    \draw [thick,\wallColor] (center) -- ++(240:1.5cm); 
    \draw [thick] (center) -- ++(0:1.5cm);   

    \drawbranchpointmarker{center};
   \node[scale=0.6] at (0.8,0) [above] {$12$};
   \draw[->] (-0.45,-0.45*1.732 ) -- (-0.451,-0.451*1.732);
      \draw[->] (-0.45,0.45*1.732 ) -- (-0.451,0.451*1.732);
         \node[scale=0.6] at (-0.45,0.45*1.732 ) [left] {$21$};
         \node[scale=0.6] at (-0.45,-0.45*1.732 ) [left] {$21$};
    \draw[->] (0.8,0 ) -- (0.9,0);
   \draw[snake it, orange] (-3, 0) to (0, 0);
    \draw [domain=0:360,line width=0.2mm] plot ({0.8*cos(\x)}, {0.8*sin(\x)});
    \node at (0,0.8) [above] {$\gamma_c$};
       \draw[<-] (0,0.8 ) -- (0.1,0.8);
%
          \draw[red,postaction={
    decorate,
    decoration={
      markings,
      mark=at position 0.5 with {\arrow{<}}
    }
  }] (-0.5*0.2,-1.732/2*0.2)  -- (-0.5*0.8,1.732/2*0.8);
                 \draw[red] (0.2,0)  -- (-0.5*0.8,1.732/2*0.8);
          \node[scale=0.6,red] at (-0.4,0.1) [above] {$2$};
      \draw [red,domain=-120:0] plot ({0.2*cos(\x)}, {0.2*sin(\x)});
   \draw [blue,domain=-120:-240] plot ({0.1*cos(\x)}, {0.1*sin(\x)});
   \draw[blue,postaction={
    decorate,
    decoration={
      markings,
      mark=at position 0.5 with {\arrow{<}}
    }
  }](-0.5*0.1,-1.732/2*0.11) -- (0.8,0);
   \draw[blue](-0.5*0.1,1.732/2*0.11) -- (0.8,0);
  \node[scale=0.6,blue] at (0.4,0.15) [above] {$1$};
%
          \draw[purple] (-0.5*0.2,1.732/2*0.2)  -- (-0.5*0.8,-1.732/2*0.8);
          \draw[purple,postaction={
    decorate,
    decoration={
      markings,
      mark=at position 0.5 with {\arrow{<}}
    }
  }] (0.2,0)  -- (-0.5*0.8,-1.732/2*0.8);
          \node[scale=0.6,purple] at (0.2,-0.5) [above] {$2$};
      \draw [purple,domain=120:0] plot ({0.2*cos(\x)}, {0.2*sin(\x)});
  \end{scope}
\end{tikzpicture}
\caption{A local picture of detours on solitons corresponding to rank 2 differential equations.}
\label{fig:Vorosconn}
\end{figure}

Thus the matrix identity now reads

\begin{equation}\label{solW2mat2}
\begin{pmatrix}1 & 1\\ 0& 1\end{pmatrix}.\begin{pmatrix}1 & 0\\ -1& 1\end{pmatrix}.\begin{pmatrix}1 & 1\\ 0& 1\end{pmatrix}.\begin{pmatrix}0 & 1\\1& 0\end{pmatrix}=\begin{pmatrix}1 & 0\\0& -1\end{pmatrix}
\end{equation}
or
\begin{equation}\label{solW2mat3}
\begin{pmatrix}1 & -1\\ 0& 1\end{pmatrix}.\begin{pmatrix}1 & 0\\ 1& 1\end{pmatrix}.\begin{pmatrix}1 & -1\\ 0& 1\end{pmatrix}.\begin{pmatrix}0 & 1\\1& 0\end{pmatrix}=\begin{pmatrix}-1 & 0\\0& 1\end{pmatrix}
\end{equation}
There is also another choice if we reverse all the $\widetilde{\gamma_{ij}}^{cb}$ and all the soliton degeneracies at the same time. We show the soliton charges for \eqref{solW2mat2} as an example in \autoref{fig:solsconn}.\footnote{We define $\gamma_c$ to be a loop on $C$ as shown in \autoref{fig:Vorosconn} and for convenience, we choose the starting point to be slightly below the branch cut. The soliton paths for the wall of type 12 and 21 should have different orientations, due to the compactibility condition that their combination should be equivalent to a trivial path. As an example, we make the choice of \autoref{fig:Vorosconn}. The diagonal matrix on the RHS  is accordingly determined by the total angle surrounding the branch points which affects \eqref{1looppart}. 
The first diagonal element is obtained from the lift of $\gamma_c$ with the starting point on sheet 1 deformed by  the wall 12 which leads to a total angle of 0 degree surrounding the branch point.  The second diagonal element is obtained from the lift of $\gamma_c$ with the starting point on sheet 2 deformed by either 21 wall or the walls 21,12,21 in a series, both lead to a total angle of $4\pi$ degrees surrounding the branch point. We see that this $4\pi$ angle leads to a minus sign in \eqref{1looppart} compared to the 0 angle which simply gives 1.}
This alternative way using soliton degeneracies and soliton charges matches the one in \autoref{mathmon}.

\subsubsection{Revisiting the nonabelianization map in the Virasoro $c=1$ case}\label{sec:matchsoln}

The setup in the construction of the nonabelianization map for the Virasoro $c=1$ algebra is in line with the convention of \autoref{sec:virsol}. However, the language is not coherent.  Using the language of \autoref{sec:virsol}, the soliton degeneracies are paired with the choice of soliton charges. Here we use \eqref{solW2mat2} as an example. We see that for wall 21, $z(q^{(+)})=z(q^{(2)})$ and $z(q^{(-)})=z(q^{(1)})$ so $z(q^{(+)})=\E^{-\pi\ri}z(q^{(-)})$,  similarly for wall 12, $z(q^{(+)})=z(q^{(1)})$ and $z(q^{(-)})=z(q^{(2)})$, so $z(q^{(+)})=\E^{\pi\ri}z(q^{(-)})$. However the soliton degeneracy for wall 21 and 12 are $1$ and $-1$ respectively. The building blocks of the change of coordinate in \cite{Hao:2024vlg} are\footnote{The details in \cite{Hao:2024vlg} is not copied in this paper. However, a similar calculation can be found for the $W_3$ case in \autoref{sec:W3-coc}.}
\begin{equation}\label{coc1}
\mu \sqrt{2z(q^{(+)})} \sqrt{2z(q^{(-)})} = 2 \I z(q^{(+)}) \text{ for $21$ wall},
\end{equation}
\begin{equation}\label{coc2}
\mu \sqrt{2z(q^{(+)})} \sqrt{2z(q^{(-)})} =-( -2 \I z(q^{(+)}) )=2 \I z(q^{(+)})\text{ for $12$ wall}.
\end{equation}
All together, the two differences of signs cancel with each other and lead to a unique expression. This is why in \cite{Hao:2024vlg},  the explicit soliton data is avoided by fixing a definite 
$z(q^{(-)})=\E^{\pi\ri}z(q^{(+)})$ for both walls
which leads to the same expression from the calculation
\begin{equation}
\sqrt{2z(q^{(+)})} \sqrt{2z(q^{(-)})} = 2 \I z(q^{(+)})
\end{equation}
which is the as \eqref{coc1} and \eqref{coc2}.
\subsection{$c=2$ $W_3$-algebra conformal blocks}\label{sec:disc_W3}
In this section, we explain the quantities appearing in \eqref{eq:EWreg-def} and we calculate them explicitly for the example of \autoref{sec:tau}.
\subsubsection{8-wall spectral network}\label{sec:8wall}
In examples of a 3-fold cover $\widetilde{C}\rightarrow C$, compared to the (ij) type of branch point which also appears in the 2-fold cover examples corresponding to Virasoro algebra, the new $(123)$ type  branch points are more chanllenging. Thus, we focus on the $(123)$ type  branch points which are described locally by $x=z^3$. Such a branch point can be obtained by the degeneration of branch points of type $(12)$ and $(23)$.  

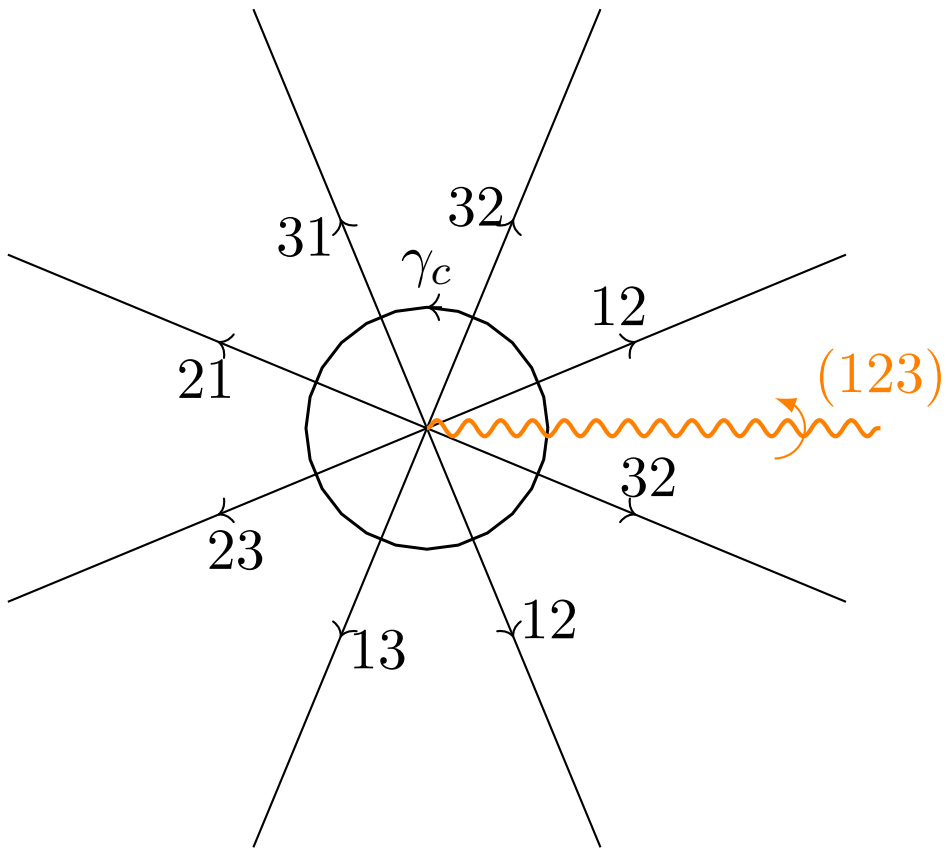
\begin{figure}
\begin{center}
\begin{tikzpicture}
  \def\labels{{"12","32","31","21","23","13","12","32"}}
  \def\r{3}
  
  \foreach \n in {0,...,7} {
    \pgfmathsetmacro{\angle}{22.5 + 45*\n}
    
    \draw[postaction={
    decorate,
    decoration={
      markings,
      mark=at position 0.5 with {\arrow{>}}
    }
  }] (0,0) -- ({\r*cos(\angle)},{\r*sin(\angle)});
    
    \node at ({(\r-1.5)*cos(\angle+10)},{(\r-1.5)*sin(\angle+10)})
      {\pgfmathparse{\labels[\n]}\pgfmathresult};
  }
  
      \draw [domain=0:360,line width=0.2mm] plot ({0.8*cos(\x)}, {0.8*sin(\x)});
    \node at (0,0.8) [above] {$\gamma_c$};
       \draw[<-] (0,0.8 ) -- (0.1,0.8);

  \draw[orange, thick, decorate, decoration={snake, amplitude=1.5pt, segment length=6pt}]
    (0,0) -- ({\r*cos(0)},{\r*sin(0)})
    node[above, sloped] {$(123)$};
    
       \draw[-latex,orange] ($(2.3,-.2)+(0,0)$) arc
    [
        start angle=90,
        end angle=-90,
        x radius=0.2cm,
        y radius =-0.2cm
    ] ;

\end{tikzpicture}
\end{center}
\caption{The 8-wall spectral network for our example.}
\label{fig:8wall}
\end{figure}

\begin{figure}
\centering
\includegraphics[width=0.5\linewidth]{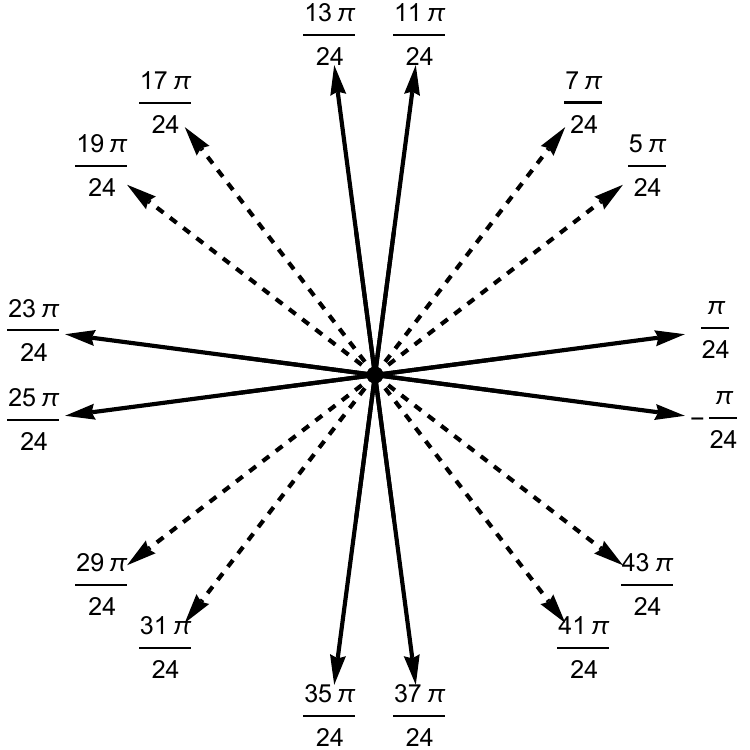}
\caption{The lift of the 8-wall spectral network to $\widetilde{C}$.}
\label{fig:16wall}
\end{figure}

In this paper, we study the simplest example with only two $(123)$ type branch points. We use the three-fold cover $\widetilde{C}=\mathbb{CP}_1$ of a Riemann surface $C=\mathbb{CP}_1$, with  the projection map 
\begin{align}
\pi:\widetilde{C}&\rightarrow C\\
z&\mapsto x=z^3.
\end{align} 
A corresponding spectral network $\cW$ is shown in \autoref{fig:8wall}\footnote{Spectral networks depend on a phase parameter $\vartheta$ and are usually denoted by $\cW^{\vartheta}$. But in this example, all of them are just rotations of the one shown in \autoref{fig:8wall}. The choice of $\vartheta$ is not important as we will see in the concrete computation. So without loss of generality, we just use $\cW$ as shown in \autoref{fig:8wall}.}. Note that this can also be viewed as the local picture of a generic spectral network around a type (123) branch point. This spectral network can be lifted to $\widetilde{C}$ as shown in \autoref{fig:16wall}. Note that in this example, $\cW$ consists of 8 walls $\cG$ which are straight lines from $0$ to $\infty$ that are spaced evenly. We can further divide the walls $\cG\in\cW$ into two groups corresponding to the label i(i+1) or i(i-1) respectively.  We call the four walls of label i(i+1) the type I or "+" and the four walls of label i(i-1) the type II or "$-$", i.e. we refer to $\cG\in\cW$ as $\cG_{I,II}$ or $\cG_{\pm}$ such that it consists of rays of angle $\alpha+\frac{\pi\ri}{2}n$, $n=0,1,2,3$  where $\alpha_-=\alpha_++\frac{\pi}{12}$. We further introduce label $\sigma=\pm 1$ to represent these two types of walls for future convenience. We then study the operator \eqref{eq:EWreg-def} which we copy here
 \begin{equation} \label{eq:EW-secsol}
E(\cW) =\exp \left[- \frac{1}{2 \pi \I} \int_\cW\mu(-\widetilde{\gamma_{ji}}^{cb}) 
\tikz[remember picture, baseline]{
  \node[anchor=base, inner sep=0pt] (psiplus) {$\psi_+(q^{(+)})^{z^{(+)}}$};
}
\, 
\tikz[remember picture, baseline]{
  \node[anchor=base, inner sep=0pt] (psiminus) {$\psi_-(q^{(-)})^{z^{(-)}}$};
}
\, \de z \right]  \, .
\end{equation}
\begin{tikzpicture}[overlay, remember picture]
  \coordinate (start) at ($(psiplus.south)!0.5!(psiplus.north)$);
  \coordinate (end) at ($(psiminus.south)!0.5!(psiminus.north)$);
  \coordinate (startShifted) at ([yshift=-2ex]start);
  \coordinate (endShifted) at ([yshift=-2ex]end);
  \draw[thick] (startShifted) -- (endShifted);
  \draw[thick] (startShifted) -- ++(0,0.5ex);
  \draw[thick] (endShifted) -- ++(0,0.5ex);
  \coordinate (mid) at ($(startShifted)!0.5!(endShifted)$);
  \node at ($(mid)+(0,-1.5ex)$) {$\ell_\cG(q)$};
\end{tikzpicture} 
Intuitively, we insert $\psi^+$ on the solid lines in \autoref{fig:16wall} and $\psi^-$ on the dashed lines in \autoref{fig:16wall}.

\subsubsection{Soliton degeneracies for \texorpdfstring{\autoref{sec:8wall}}{Chapter 5}}\label{sec:solitonW3}

We need to first calculate the soliton degeneracies. Note that type I walls naturally come with a counterclockwise orientation soliton charge and similarly type II walls come with a clockwise orientation. The calculation of soliton degeneracies boils down to the following matrix equality:\footnote{The RHS is an identity matrix because of the following. Analogous to \eqref{1looppart} for the Virasoro case, there is a phase problem for a loop $\gamma$ around the branch point in the $W_3$-algebra case. It is determined by 
\begin{equation}
\label{monW3} \E^{-\frac{1}{3} \oint\limits_\gamma \de \log P }.
\end{equation}
We define $\gamma_c$ to be a loop on $C$ as shown in \autoref{fig:8wall} and for convenience, we choose the starting point to be slightly above the branch cut. The first diagonal element is obtained from the lift of $\gamma_c$ with the starting point on sheet 1 deformed by both walls of type 12 and 23 or the lift of $\gamma_c$ with the starting point on sheet one deformed by either wall of type 13. They both give a total $6\pi$ angle winding around the branch point; the second diagonal element is obtained from the lift of $\gamma_c$  with the starting point on sheet 2 deformed by wall 21, which corresponds to total angle 0; and the third diagonal element comes from the lift of $\gamma_c$ with the starting point on sheet 3 deformed by  a wall 32, or the lift of $\gamma_c$ with the starting point on sheet 3 deformed by walls 32, 23 and 32 in a series or the lift of $\gamma_c$ with the starting point on sheet 3 deformed by walls 31, 13 and 32,  or the lift of $\gamma_c$ with the starting point on sheet 3 deformed by walls 32,  21, 13 and 32 in a series whose total angles all equal to 0. We see that both 0 and $6\pi$ angles lead to a trivial 1 in \eqref{monW3}.}
\begin{align}
\nonumber&\begin{pmatrix}
1 & 0 & 0\\
1 & 1 & 0\\
0 & 0 & 1
\end{pmatrix}
\cdot
\begin{pmatrix}
1 & 0 & 0\\
0 & 1 & -1\\
0 & 0 & 1
\end{pmatrix}
\cdot
\begin{pmatrix}
1 & 0 & 1\\
0 & 1 & 0\\
0 & 0 & 1
\end{pmatrix}
\cdot
\begin{pmatrix}
1 & -1 & 0\\
0 & 1 & 0\\
0 & 0 & 1
\end{pmatrix}
\cdot
\begin{pmatrix}
1 & 0 & 0\\
0 & 1 & 0\\
0 & -1 & 1
\end{pmatrix}
\cdot
\begin{pmatrix}
1 & 0 & 0\\
0 & 1 & 0\\
-1 & 0 & 1
\end{pmatrix}
\cdot\\
&
\cdot\begin{pmatrix}
1 & 0 & 0\\
1 & 1 & 0\\
0 & 0 & 1
\end{pmatrix}
\cdot
\begin{pmatrix}
1 & 0 & 0\\
0 & 1 & -1\\
0 & 0 & 1
\end{pmatrix}
\cdot
\begin{pmatrix}
0 & 1 & 0\\
0 & 0 & 1\\
1 & 0 & 0
\end{pmatrix}
=\left(
\begin{array}{ccc}
 1 & 0 & 0 \\
 0 & 1 & 0 \\
 0 & 0 & 1 \\
\end{array}
\right)\,\,.
\end{align}

We get 
\begin{align}
\mu(-\widetilde{\gamma_{(i+1)i}}^{cb}) &=1,\\
\mu(-\widetilde{\gamma_{i(i+1)}}^{cb}) &=-1.
\end{align}
\subsubsection{Nonabelianization map for \texorpdfstring{\autoref{sec:tau}}{Chapter 5}}\label{sec:W3-coc}
Following the convention of the two types of walls, in the example that we study in \autoref{sec:tau}, \eqref{eq:EW-secsol} can be written as 
\begin{align}
\nonumber E(\cW)& = \exp \left[ -\frac{1}{2 \pi \I} \int_\cW \mu(-\widetilde{\gamma_{ji}}^{cb}) \, \tikz[remember picture, baseline]{
  \node[anchor=base, inner sep=0pt] (psiplus) {$\psi_+(q^{(+)})^{z^{(+)}}$};
}
\, 
\tikz[remember picture, baseline]{
  \node[anchor=base, inner sep=0pt] (psiminus) {$\psi_-(q^{(-)})^{z^{(-)}}$};
}
 \, \de z \right] \\
\label{sol}& = \exp \left[ -\frac{1}{2 \pi \I} \int_{\cG_+} \tikz[remember picture, baseline]{
  \node[anchor=base, inner sep=0pt] (psiplus2) {$\psi_+(q^{(+)})^{z^{(+)}}$};
}
\, 
\tikz[remember picture, baseline]{
  \node[anchor=base, inner sep=0pt] (psiminus2) {$\psi_-(q^{(-)})^{z^{(-)}}$};
}
\, \de z +\frac{1}{2 \pi \I} \int_{\cG_{-}} \tikz[remember picture, baseline]{
  \node[anchor=base, inner sep=0pt] (psiplus3) {$\psi_+(q^{(+)})^{z^{(+)}}$};
}
\, 
\tikz[remember picture, baseline]{
  \node[anchor=base, inner sep=0pt] (psiminus3) {$\psi_-(q^{(-)})^{z^{(-)}}$};
}
 \, \de z \right] \, .
 \end{align}
 \begin{tikzpicture}[overlay, remember picture]
  \coordinate (start) at ($(psiplus.south)!0.5!(psiplus.north)$);
  \coordinate (end) at ($(psiminus.south)!0.5!(psiminus.north)$);
  \coordinate (startShifted) at ([yshift=-2ex]start);
  \coordinate (endShifted) at ([yshift=-2ex]end);
  \draw[thick] (startShifted) -- (endShifted);
  \draw[thick] (startShifted) -- ++(0,0.5ex);
  \draw[thick] (endShifted) -- ++(0,0.5ex);
  \coordinate (mid) at ($(startShifted)!0.5!(endShifted)$);
  \node at ($(mid)+(0,-1.5ex)$) {$\ell_\cG(q)$};
\end{tikzpicture} 
\begin{tikzpicture}[overlay, remember picture]
  \coordinate (start) at ($(psiplus2.south)!0.5!(psiplus2.north)$);
  \coordinate (end) at ($(psiminus2.south)!0.5!(psiminus2.north)$);
  \coordinate (startShifted) at ([yshift=-2ex]start);
  \coordinate (endShifted) at ([yshift=-2ex]end);
  \draw[thick] (startShifted) -- (endShifted);
  \draw[thick] (startShifted) -- ++(0,0.5ex);
  \draw[thick] (endShifted) -- ++(0,0.5ex);
  \coordinate (mid) at ($(startShifted)!0.5!(endShifted)$);
  \node at ($(mid)+(0,-1.5ex)$) {$\ell_\cG(q)$};
\end{tikzpicture} 
\begin{tikzpicture}[overlay, remember picture]
  \coordinate (start) at ($(psiplus3.south)!0.5!(psiplus3.north)$);
  \coordinate (end) at ($(psiminus3.south)!0.5!(psiminus3.north)$);
  \coordinate (startShifted) at ([yshift=-2ex]start);
  \coordinate (endShifted) at ([yshift=-2ex]end);
  \draw[thick] (startShifted) -- (endShifted);
  \draw[thick] (startShifted) -- ++(0,0.5ex);
  \draw[thick] (endShifted) -- ++(0,0.5ex);
  \coordinate (mid) at ($(startShifted)!0.5!(endShifted)$);
  \node at ($(mid)+(0,-1.5ex)$) {$\ell_\cG(q)$};
\end{tikzpicture} 

Technically, the appearance of two 
different local coordinate systems
$x^{(\pm)}$ on $\tC$ is inconvenient, and note that $\ell_\cG(q)$ carries 
the spin structure $\pi^* K_C^{\frac12}$, which is not globally
defined on $\tC$. We need to rewrite \eqref{sol}.
Let $\ell'_\cG(q)$ be the same path but now with the
standard spin structure $K_\tC^{\frac12}$ on $\tC = \bbC\bbP^1$, 
and fix an 
isomorphism $\iota: \pi^* K_C^{\frac12} \to K^{\frac12}_\tC$ along the path $\ell_\cG(q)$.
We
rewrite the exponent of \eqref{eq:EW-secsol} as
\Needspace{5\baselineskip}
\begin{equation} \label{eq:wall-basic-rewritten}
  \int_\cG \de x(q) \, \left(
   \tikz[remember picture, baseline]{
  \node[anchor=base, inner sep=0pt] (psiplus) {$\psi_+(q^{(+)})^z$};
}     \frac{\sqrt{\de z(q^{(+)})}}{\iota(\pi^*\sqrt{\de x(q)})} \right)
    \left(
    \tikz[remember picture, baseline]{
  \node[anchor=base, inner sep=0pt] (psiminus) {$\psi_-(q^{(-)})^z$};
}    \frac{\sqrt{\de z(q^{(-)})}}{\iota(\pi^*\sqrt{\de x(q)})} \right) \, .
\end{equation}
\begin{tikzpicture}[overlay, remember picture]
  \coordinate (start) at ($(psiplus.south)!0.5!(psiplus.north)$);
  \coordinate (end) at ($(psiminus.south)!0.5!(psiminus.north)$);
  \coordinate (startShifted) at ([yshift=-4ex]start);
  \coordinate (endShifted) at ([yshift=-4ex]end);
  \draw[thick] (startShifted) -- (endShifted);
  \draw[thick] (startShifted) -- ++(0,0.5ex);
  \draw[thick] (endShifted) -- ++(0,0.5ex);
  \coordinate (mid) at ($(startShifted)!0.5!(endShifted)$);
  \node at ($(mid)+(0,-1.5ex)$) {$\ell_\cG(q)$};
\end{tikzpicture} 

{\hspace{1cm}}

To simplify this, note that 
$x(q) = z(q^{(+)})^3$, so $\de x(q) = 3z(q^{(+)})^2 \, \de z(q^{(+)})$. It follows that
$\iota(\pi^* \sqrt{\de x(q)}) = \sqrt{3}z(q^{(\pm)}) \sqrt{\de z(q^{(\pm)})}$, for type I and II walls $\sqrt{3}z(q^{(+)}) \sqrt{3}z(q^{(-)}) = 3 \E^{\pm\frac{2\pi\ri}{3}} z(q^{(+)})^2$.
Thus the insertion can be simplified to
\begin{equation}
  \int\limits_{\cG_{\sigma}} \frac{3z(q^{(+)})^2 \, \de z(q^{(+)})}{3 \E^{\pm\frac{2\pi\ri}{3}} z(q^{(+)})^2} \tikz[remember picture, baseline]{
  \node[anchor=base, inner sep=0pt] (psiplus) {$\psi_+(q^{(+)})^z$};
} \tikz[remember picture, baseline]{
  \node[anchor=base, inner sep=0pt] (psiminus) {$\psi_-(q^{(-)})^z$};
}    =  \E^{\mp\frac{2\pi\ri}{3}}  \int_{\cG_{\sigma}}  \de z(q^{(+)})
\tikz[remember picture, baseline]{
  \node[anchor=base, inner sep=0pt] (psiplus2) {$\psi_+(q^{(+)})^z$};
} \tikz[remember picture, baseline]{
  \node[anchor=base, inner sep=0pt] (psiminus2) {$\psi_-(q^{(-)})^z$};
}   \, .
\end{equation}
 \begin{tikzpicture}[overlay, remember picture]
  \coordinate (start) at ($(psiplus.south)!0.5!(psiplus.north)$);
  \coordinate (end) at ($(psiminus.south)!0.5!(psiminus.north)$);
  \coordinate (startShifted) at ([yshift=-2ex]start);
  \coordinate (endShifted) at ([yshift=-2ex]end);
  \draw[thick] (startShifted) -- (endShifted);
  \draw[thick] (startShifted) -- ++(0,0.5ex);
  \draw[thick] (endShifted) -- ++(0,0.5ex);
  \coordinate (mid) at ($(startShifted)!0.5!(endShifted)$);
  \node at ($(mid)+(0,-1.5ex)$) {$\ell_\cG(q)$};
\end{tikzpicture} 
\begin{tikzpicture}[overlay, remember picture]
  \coordinate (start) at ($(psiplus2.south)!0.5!(psiplus2.north)$);
  \coordinate (end) at ($(psiminus2.south)!0.5!(psiminus2.north)$);
  \coordinate (startShifted) at ([yshift=-2ex]start);
  \coordinate (endShifted) at ([yshift=-2ex]end);
  \draw[thick] (startShifted) -- (endShifted);
  \draw[thick] (startShifted) -- ++(0,0.5ex);
  \draw[thick] (endShifted) -- ++(0,0.5ex);
  \coordinate (mid) at ($(startShifted)!0.5!(endShifted)$);
  \node at ($(mid)+(0,-1.5ex)$) {$\ell_\cG(q)$};
\end{tikzpicture} 
Finally we simplify our notation as follows. We 
write just $z$ for $z(q^{(+)})$, and use always the standard
coordinate $z$, the standard spin structure $K_\tC^{\frac12}$,
and the standard $\sqrt{\de z}$ on $\tC = \bbC\bbP^1$. The leash in this example is determined uniquely . For simplicity, we also ignore the leash label. In the calculation in \autoref{sec:tau}, we consider products of
\begin{equation}\label{eq:pluswallint}
-\left(\frac{1}{2\pi\ri}\right) \e^{-\frac{2\pi\ri}{3}}\int\limits_{\cG_+} \de z
 \psi_+(z)\psi_-(z\e^{\frac{2\pi\ri}{3}}) 
\end{equation}
and
\begin{equation}\label{eq:minuswallint}
+\left(\frac{1}{2\pi\ri}\right) \e^{\frac{2\pi\ri}{3}}\int\limits_{\cG_{-}} \de z
 \psi_+(z)\psi_-(z\e^{-\frac{2\pi\ri}{3}}).
\end{equation}
appearing in the expansion of \eqref{eq:EW-secsol}. For convenience, we use the $\sigma=\pm 1$ notation introduced before to represent the type of walls and define
\begin{equation}
\Theta_{\sigma}=\E^{\frac{2\pi\ri}{3}\sigma}.
\end{equation}
Then \eqref{eq:pluswallint} and \eqref{eq:minuswallint} can be written uniformly as
\begin{equation}
-\left(\frac{1}{2\pi\ri}\right)\E^{-\frac{\pi\ri}{2}+\frac{\pi\ri\sigma}{2}} \Theta_{(-\sigma)}\int\limits_{\cG_\sigma} \de z
 \psi_+(z)\psi_-(\Theta_{\sigma}z). 
\end{equation}

\subsubsection{Some numerical checks for the discontinuities}\label{sec:checkdisc}

The result above matches
the discontinuities that we expected. Similar to the Virasoro case, we have
\begin{align}
\nonumber&\quad\disc_{w\rightarrow \cG^{(+)}}\left(\psi_+(w)\psi_-(v) \left(- \frac{1}{2\pi\ri}\right)\e^{-\frac{2\pi\ri}{3}}\int\limits_{\cG_+} \de z\psi_+(z)\psi_-(z\e^{\frac{2\pi\ri}{3}}) \right)\\
\label{jump1}&=-\e^{\frac{2\pi\ri}{3}}\psi_+(w\e^{-\frac{2\pi\ri}{3}})\psi_-(v)
\end{align}

and similarly
\begin{align}
\nonumber&\quad\disc_{w\rightarrow \cG^{(-)}}\left(\psi_+(w)\psi_-(v)  \left(\frac{1}{2\pi\ri}\right) \e^{\frac{2\pi\ri}{3}}\int\limits_{\cG_{-}} \de z
 \psi_+(z)\psi_-(z\e^{-\frac{2\pi\ri}{3}})\right)\\
\label{jump2}&=\e^{-\frac{2\pi\ri}{3}}\psi_+(w\e^{\frac{2\pi\ri}{3}})\psi_-(v)
\end{align}
where the discontinuity is defined as $w$ in the counterclockwise direction of the wall minus the clockwise direction. Note that here $w,z$ are coordinates on $\tC $. Let $x=w^3$ be the coordinate on $C$. If we view $\psi_+(w)$ as $\psi^{(1)}(x) $,  $\psi_+(w\e^{-\frac{2\pi\ri}{3}})$ as $\psi^{(2)}(x) $ and $\psi_+(w\e^{\frac{2\pi\ri}{3}})$ as $\psi^{(3)}(x) $, the coefficients are as expected for the NS phase Stokes multipliers. In \autoref{solcheckW3}, we show a numerical check of the Stokes multipliers or soliton data in the generalized Airy example whose spectral network is the same as \autoref{fig:8wall} except that the two branch points are also singular points. But this doesn't affect the  calculation of soliton degeneracies. The Airy example corresponds to a third order differential equation
\begin{equation}\label{eqn:genAiry}
-\hbar^3\Psi^{(3)}(x)+x\Psi(x)=0.
\end{equation}
And the all-order WKB solutions read
 \begin{equation}\label{gAirysol}
\Psi_{i}(x)=\E^{\frac{1}{\hbar}\int_{\infty}^x \sum_{n>0}\lambda_n^{l}(x)\hbar^n\rd x},\text{ where }\lambda_0^{l}(x)=\E^{-(l-1)\frac{2\pi\ri}{3}}x^{\frac13}.
\end{equation} 
In this example, using $x=w^3$ as above,
 \begin{align}
 \frac{1}{\hbar}\int_{\infty}^x \sum_{n>0}\lambda_n^{l}(x)\hbar^n\rd x&\sim x^{-\frac{4}{3} (n-1)} \hbar ^{n-1} \E^{l\frac{2\pi\ri}{3}(n-1)}= (w\E^{-l\frac{2\pi\ri}{3}})^{-4 (n-1)} \hbar ^{n-1}, &\text{ for }n\neq 1,\\
 \frac{1}{\hbar}\int_{\infty}^x \sum_{n>0}\lambda_n^{l}(x)\hbar^n\rd x&=-\frac{\log x}{3},&\text{ for }n= 1.
 \end{align}
In particular, we take $\hbar=1$ which corresponds to the $c=2$ case we study in this paper. Note that $\Psi_{l}(x) =\Psi_1(w\e^{-\frac{2\pi\ri}{3}(l-1)})$ as expected. We show the discontinuities of  $s(\Psi_1)(\epsilon^{\pm})$ in \autoref{solcheckW3} where $\epsilon^+$ is a larger and  $\epsilon^-$ is a smaller perturbation of the critical phase. Since the walls rotate counterclockwise, we see opposite signs to \eqref{jump1} and \eqref{jump2}. 

In fact the exact WKB of 4d $N=2$ $SU(N)$ theories is a much broader field attracting more attentions. More discussions related to the spectral network and exact WKB can be found in \cite{Gaiotto:2009hg,Gaiotto:2012rg,Gaiotto:2014bza,Yan:2020kkb,Kineider:2024ttx,Hao:2025azt,Nemes2025Borel}.
%

\begin{table}[htp]
\caption{Numerical check for discontinuities of the solution $\Psi_1$ of \eqref{gAirysol} at $x=3$.}
\begin{center}
\begin{tabular}{|c|c|c|c|}
\hline
$\epsilon$ & $s(\Psi_1)(\epsilon^+)-s(\Psi_1)(\epsilon^-)$ & $\E^{\frac{2\pi\ri}{3}}s(\Psi_2)(\epsilon)$ & $-\E^{-\frac{2\pi\ri}{3}}s(\Psi_3)(\epsilon)$\\
\hline
$\frac{1}{3}\E^{\frac{\pi}{6}\ri}$ & $-0.000139-0.000055\ri$ & $-0.00013889+0.00005471\ri$ & \\
\hline
$\frac{1}{3}\E^{-\frac{\pi}{6}\ri}$ & $0.000139-0.000055\ri$ & &$0.00013889-0.00005471\ri$  \\
\hline
\end{tabular}
\end{center}
\label{solcheckW3}
\end{table}%

\section{A simple example}\label{sec:tau}

In this section, we show an explicit calculation for the simple example of the three-fold cover $\widetilde{C}=\mathbb{CP}_1$ of a Riemann surface $C=\mathbb{CP}_1$, with  the projection map 
\begin{align}
\pi:\widetilde{C}&\rightarrow C\\
z&\mapsto x=z^3.
\end{align} 
The corresponding spectral networks and soliton degeneracies have been shown in \autoref{sec:8wall} and \autoref{sec:solitonW3}. In this case, the space of conformal blocks is one dimensional and we denote a conformal block by $\tPsi$. 
Moreover, the leash notation in this example is ignored as before.

\subsection{Two-point correlator}\label{tau:2point}
We start by calculating
\begin{equation}
F(z,w) = \frac{ \langle
\tikz[remember picture, baseline]{
  \node[anchor=base, inner sep=0pt] (psiplus) {$\psi_+(z)$};
}
\,
\tikz[remember picture, baseline]{
  \node[anchor=base, inner sep=0pt] (psiminus) {$\psi_-(w)$};
}
\, E_\ren(\cW) \rangle_\tPsi }{ \langle E_\ren(\cW) \rangle_\tPsi} =\sum_{n=0}^\infty\sum_{\{\sigma_1,\cdots,\sigma_n\}}w_{\sigma_1\cdots \sigma_n}\, ,
\end{equation}
where
\begin{equation}
w_{\sigma_1\cdots \sigma_n}=\E^{\ri\pi n}\prod_{i=1}^n\Bigg(-\left(\frac{1}{2\pi\ri}\right)\E^{-\frac{\pi\ri}{2}+\frac{\pi\ri\sigma_i}{2}} \Theta_{(-\sigma_i)}\int\limits_{\cG_{\sigma_i}} 
 \psi_+(z)\psi_-(\Theta_{\sigma_i}z)  \,\de z\Bigg)  \, .
\end{equation}
Recall that $\cG_\sigma$ consists of rays of the angles $\alpha_\sigma+\frac{\pi\ri}{2}n$, $n=0,1,2,3$ --- $\cG_+$ is along the directions $\alpha_++n\frac{\pi\ri}{2}$ and $\cG_{-}$ is along the directions $\alpha_-+n\frac{\pi\ri}{2}$, where $\alpha_-=\alpha_++\frac{\pi}{12}$\footnote{In the generic expression of terms below, we first treat them as independent and use them symbolicly. We will see later that the dependence on $\alpha_{\pm}$ all cancel.}. Without loss of generality, define $\alpha_{\pm 1}\equiv\alpha_{\pm}={\mp}\frac{\pi}{24}$. And as a quick reminder $\Theta_{\pm 1}\equiv\Theta_{\pm}=\e^{\pm\frac{2\pi\ri}{3}}$. 
First several terms can be found in \autoref{appen:Fzw-exp} where we used \eqref{1wallsumtrickgen} and \eqref{1wallsumtrick}. And the generic expression can be rewritten as %
\begin{equation}
\begin{aligned}
w_{\sigma_1\cdots\sigma_n}=&\E^{-\frac{n\pi\ri}{2}+\frac{\pi}{2}\ri\sum\limits_i\sigma_i }\left(\frac{1}{2\pi\ri}\right)^n(\prod\limits_{i=1}^n\Theta_{(-\sigma_i)})\Theta_{(-\sigma_1)}\left(w+(\prod\limits_{i=1}^n\Theta_{(-\sigma_i)}z)\right)\left(w^2+(\prod\limits_{i=1}^n \Theta_{(-\sigma_i)}z)^2\right)\times\\
&\left(\prod\limits_{i=1}^n\,\int\limits_{\mathbb{R}_+}\rd t_i\right)  \frac{ 1}{\left( (\Theta_{-\sigma_1} z)^4-\e^{4 \ri \alpha_{\sigma_1}} t_1\right)}\left(\prod\limits_{i=2}^{n}\frac{\e^{4 \ri \alpha_{\sigma_{i-1}}} }{\left(\e^{4 \ri \alpha_{\sigma_{i-1}}} t_{i-1}-\e^{4 \ri \alpha_{\sigma_i}} \Theta_{{\sigma_i}}t_i\right)}\right)\frac{\e^{4 \ri \alpha_{\sigma_n}} }{\left(\e^{4 \ri \alpha_{\sigma_n}} t_n-w^4\right)}\\
=&\E^{-\frac{n\pi\ri}{2}+\frac{\pi}{2}\ri\sum\limits_i\sigma_i }\left(\frac{1}{2\pi\ri}\right)^n(\prod\limits_{i=1}^n\Theta_{(-\sigma_i)})\left(w+(\prod\limits_{i=1}^n\Theta_{(-\sigma_i)}z)\right)\left(w^2+(\prod\limits_{i=1}^n \Theta_{(-\sigma_i)}z)^2\right)\times\\
&\left(\prod\limits_{i=1}^n\,\int\limits_{\mathbb{R}_+}\rd t_i\right)  \frac{ 1}{\left(  z^4-\e^{4 \ri \alpha_{\sigma_1}}\Theta_{\sigma_1} t_1\right)}\left(\prod\limits_{i=2}^{n}\frac{1 }{\left( t_{i-1}-\e^{4 \ri \left(\alpha_{\sigma_i}-\alpha_{\sigma_{i-1}}\right)} \Theta_{{\sigma_i}}t_i\right)}\right)\frac{1 }{\left( t_n-\e^{-4 \ri \alpha_{\sigma_n}}w^4\right)},
\end{aligned}
\end{equation}
where $\sigma_i=\pm 1$.
We see that the basic ingredient of the integrals are of the form
\begin{equation}
\left(\frac{1}{2\pi\ri}\right)\int \frac{f(t_{i-1})}{t_{i-1}-\E^{4\ri \Delta_i}\Theta_{\sigma_i}t_{i}}\rd t_{i-1},
\end{equation}
where
\begin{equation}
\Delta_i=\alpha_{\sigma_i}-\alpha_{\sigma_{i-1}}.
\end{equation}
We thus define the integral operator $\mathcal{I}$ such that 
\begin{equation}
\mathcal{I}_{\sigma_{i-1}}f(t_{i-1})=\left(\frac{1}{2\pi\ri}\right)\int \frac{f(t_{i-1})}{t_{i-1}-\E^{4\ri \Delta_i}\Theta_{\sigma_i}t_{i}}\rd t_{i-1}.
\end{equation}
An eigenfunction of $\mathcal{I}_{\sigma_{i-1}}$ for arbitrary $i$ is 
\begin{equation}
\phi_\lambda(t_{i-1})=t_{i-1}^{-\frac12-\ri\lambda}= \e^{(-\frac12 - \I \lambda)\log t_{i-1}},
\end{equation}
and the eigenvalue is given by
\begin{align}
\mathcal{I}_{\sigma_{i-1}}\phi_\lambda(t_{i-1})=&\left(\frac{1}{2\pi\ri}\right)\int \frac{\phi_\lambda(t_{i-1})}{t_{i-1}-\E^{4\ri \Delta_i}\Theta_{\sigma_i}t_{i}}\rd t_{i-1}\\
=&\E^{-\frac{\pi\ri}{2}}\frac12\sech(\pi\lambda)\E^{-\lambda\frac{\pi}{3}\sigma_i}\E^{\frac{\pi\ri}{6}{\sigma_i}}(\E^{4\ri\Delta_i})^{-\frac12-\ri\lambda}\phi_\lambda(t_{i}).
\end{align}
Any $f \in L^2(\bbR^+)$ can be expanded in $\phi_{\lambda}$ using
\begin{equation}
f(x) = \int_{-\infty}^\infty \hat f(\lambda)\phi_{\lambda}(x) \, \de \lambda \, ,\qquad 
\hat f(\lambda) = \frac{1}{2\pi}\int_{0}^\infty \phi_{-\lambda}(x) f(x) \, \de x \, .
\end{equation}
Now we regard
\begin{equation}
f_0(t_1)= \frac{ 1}{\left(  z^4-\e^{4 \ri \alpha_{\sigma_1}}\Theta_{\sigma_1} t_1\right)}
\end{equation}
as the input function. In particular, we can expand $f_0$ in the basis of eigenfunctions: we have
\begin{equation}
  \hat f_0(\lambda) = \frac{1}{2\pi}\int_{0}^\infty \phi_{-\lambda}(t_1) f_0(t_1) \, \de t_1 =\frac{1}2 \text{sech}(\pi\lambda)\E^{-\lambda\frac{\pi}{3}\sigma_1}\E^{\frac{\pi\ri}{6}\sigma_1}(\e^{4 \ri \alpha_{\sigma_1}})^{-\frac12-\ri\lambda} \phi_{-\lambda}(z^4) \, .  
\end{equation}
Thus 
\begin{equation}\label{zwchecktheta}
\begin{aligned}
w_{\sigma_1\cdots\sigma_n}=&\E^{-\frac{n\pi\ri}{2}+\frac{\pi}{2}\ri\sum\limits_i\sigma_i }\left(w+(\prod\limits_{i=1}^n\Theta_{(-\sigma_i)}z)\right)\left(w^2+(\prod\limits_{i=1}^n \Theta_{(-\sigma_i)}z)^2\right)\times\\
&\left(\prod\limits_{i=1}^n\,\frac{1}{2\pi\ri}\int\limits_{\mathbb{R}_+}\rd t_i\right)  \frac{ \Theta_{(-\sigma_1)}}{\left(  z^4-\e^{4 \ri \alpha_{\sigma_1}}\Theta_{\sigma_1} t_1\right)}\left(\prod\limits_{i=2}^{n}\frac{\Theta_{(-\sigma_i)} }{\left( t_{i-1}-\e^{4 \ri \left(\alpha_{\sigma_i}-\alpha_{\sigma_{i-1}}\right)} \Theta_{{\sigma_i}}t_i\right)}\right)\frac{1 }{\left( t_n-\e^{-4 \ri \alpha_{\sigma_n}}w^4\right)}\\
=&\E^{\pi\ri n}\left(\prod\limits_{i=1}^n\Theta_{\sigma_i}w+z\right)\left((\prod\limits_{i=1}^n\Theta_{\sigma_i}w)^2+z^2\right)\times\\
&\int_{-\infty}^\infty \left(\frac{1}{2}\sech(\pi\lambda)\right)^{n+1}\phi_{-\lambda}(z^4) \phi_{\lambda}(-w^4)\E^{-\ri\lambda(\sum\limits_{i=1}^n\log(-\Theta_{\sigma_i}))}\rd \lambda .
\end{aligned}
\end{equation}

Notice that in \eqref{zwchecktheta}, we need to sum over the partition $\{\sigma_1,\cdots,\sigma_n\}$ for each $n$. Therefore, it is more convenient to rewrite
\begin{equation}
\begin{aligned}
w_{\sigma_1\cdots\sigma_n}=&\E^{\pi\ri n}\left(\E^{\frac{2\pi \ri}{3}(\sum\limits_{i=1}^n\sigma_i)}w+z\right)\left((\E^{\frac{2\pi \ri}{3}(\sum\limits_{i=1}^n\sigma_i)}w)^2+z^2\right)\times\\
&\int_{-\infty}^\infty \left(\frac{1}{2}\sech(\pi\lambda)\right)^{n+1}\phi_{-\lambda}(z^4) \phi_{\lambda}(-w^4)\E^{-\frac\pi 3\lambda(\sum\limits_{i=1}^n\sigma_i)}\rd \lambda
\end{aligned}
\end{equation}
in terms of 
\begin{equation}\label{sumsigma}
\sum\limits_{i=1}^n\sigma_i=S_k\equiv(+1)\times k+(-1)\times(n-k)=2k-n
\end{equation}
with multiplicity 
\begin{equation}\label{multsigma}
n_{S_k}=\binom{n}{k}.
\end{equation}
Thus
\begin{align}\label{Fzwint}
\nonumber F(z,w)&=\quad \sum_{n=0}^\infty\sum_{\{\sigma_1,\cdots,\sigma_n\}} w_{\sigma_1\cdots \sigma_n} \\
\nonumber&=\sum_{n=0}^\infty\sum_{k=0}^n \binom{n}{k} \E^{\ri\pi n}\left(z+w\E^{\frac{2\pi \ri}{3}(2k-n)}\right)\left(z^2+w^2\E^{\frac{4\pi \ri}{3}(2k-n)}\right)\times\\
\nonumber&\quad\quad\times\int_{-\infty}^\infty \left(\frac{1}{2}\sech(\pi\lambda)\right)^{n+1}\E^{-\frac\pi 3\lambda(2k-n)} \e^{(-\frac12 + \I \lambda)\log (z^4)}\e^{(-\frac12 - \I \lambda)\log (-w^4)}
\rd \lambda\, \\
\nonumber&=\int_{-\infty}^\infty\frac{\text{sech}(\pi  \lambda ) }{2 \left(\E^{4 \pi  \lambda
   }+1\right)}\E^{ \left(2 \pi  \lambda +\ri \lambda  \left(\log
   \left(z^4\right)-\log \left(-w^4\right)\right)+\frac{1}{2} \left(-\log
   \left(-w^4\right)-\log \left(z^4\right)\right)\right) }\times\\
 \nonumber  &\quad\quad\times\Bigg(2 \cosh (2 \pi  \lambda
   ) (w+z) \left(w^2+z^2\right)-\cosh \left(\frac{4 \pi  \lambda }{3}\right) (w+z)
   \left(2 w^2-3 w z+2 z^2\right)\\
   &\quad\quad+(w+z) \left(w^2+z^2\right)-\ri \sqrt{3} w z \sinh
   \left(\frac{4 \pi  \lambda }{3}\right) (w-z)\Bigg)\rd \lambda\, .
\end{align}
This integral can be calculated by closing the integral contour in either the upper or
the lower half-plane (when $\text{Re}\Big(\left(\log
   \left(z^4\right)-\log \left(-w^4\right)\right)\Big)>0$ or $\text{Re}\Big(\left(\log
   \left(z^4\right)-\log \left(-w^4\right)\right)\Big)<0$) and then using residues. For example, let's look at the part
   \begin{equation}
P_1=\frac{\text{sech}(\pi  \lambda ) (w+z) \left(w^2+z^2\right) \E^{ \left(2 \pi  \lambda
   +i \lambda  \left(\log \left(z^4\right)-\log \left(-w^4\right)\right)+\frac{1}{2}
   \left(-\log \left(-w^4\right)-\log \left(z^4\right)\right)\right)}}{2 \left(\E^{4 \pi 
   \lambda }+1\right)}.   
   \end{equation} 
Assume $\text{Re}\Big(\left(\log
   \left(z^4\right)-\log \left(-w^4\right)\right)\Big)>0$, so we only need to consider poles in the upper half plane. $P_1$ has poles at
 \begin{align}
 p_{1,1,n}&=\frac{\ri}{4}+\ri n,\\
 p_{1,2,n}&=\frac{\ri}{2}+\ri n,\\
 p_{1,3,n}&=\frac{3\ri}{4}+\ri n,
 \end{align}
 with residues
 \begin{align}
 \text{Res}_{p_{1,1,n}}&=-\frac{\ri (-1)^{n} (w+z) \left(w^2+z^2\right)
   \left(-\frac{w^4}{z^4}\right)^{n}}{4 \sqrt{2} \pi  \sqrt[4]{-w^4}
   \left(z^4\right)^{3/4}},\\
  \text{Res}_{p_{1,2,n}}&=\frac{\ri (-1)^{n} (w+z) \left(w^2+z^2\right)
   \left(-\frac{w^4}{z^4}\right)^{n}}{4 \pi  z^4},\\
  \text{Res}_{p_{1,3,n}}&=-\frac{\ri (-1)^{n} \sqrt[4]{-w^4} (w+z) \left(w^2+z^2\right)
   \left(-\frac{w^4}{z^4}\right)^{n}}{4 \sqrt{2} \pi  \left(z^4\right)^{5/4}}.
 \end{align}
Similarly, we can get all the poles and residues for other terms. We then apply the residue theorem to compute \eqref{Fzwint}. $F(z,w)$ is piecewise analytic where the discontinuities in $z$ is discussed in \autoref{sec:checkdisc} and \autoref{sec:solitonW3} and the discontinuities in $w$ can be obtained analogously. When $z>0,w>0$,
\begin{equation}\label{eq:two-point-with-walls}
F(z,w)=3zw\frac1{z^3-w^3}.
\end{equation}
Result for other values of $z,w$ can be obtained similarly. In fact, the dependence of \eqref{eq:two-point-with-walls} on $z$ and $w$ is what we expect from \eqref{eq:chi-dictionary}. Every insertion of a primary field $V_{\pm}(p^{(i)})^{z^{(i)}}$ on the cover corresponds to insertions of $W_3\otimes \Heis$ primaries $W_{\frac{1}{3},\pm\frac{\sqrt{2}}{9\sqrt{3}}}(p)^xV_{\pm\frac{1}{\sqrt{3}}}(p)^x$ on the base where the projection map is given by $x=z^3$. Thus the dependence on $z$ and $w$ should always be in the form of $z^3$ and $w^3$ up to an overall factor coming from the change of coordinates with $\frac{d x}{d z}=3 z^2$. $W_{\frac{1}{3},\pm\frac{\sqrt{2}}{9\sqrt{3}}}$ has conformal weight $\frac{1}{3}$ and $V_{\pm\frac{1}{\sqrt{3}}}$ has conformal weight $\frac12(\pm\frac{1}{\sqrt{3}})^2=\frac16 $. For simplicity, let's assume $z,w>0$. Thus under coordinate transform we get $(\frac{d x}{d z})^{\frac{1}{3}+\frac16 }=(3)^{\frac{1}{2}} z $ and similarly for $w$ we get $(3)^{\frac{1}{2}} w $. All together, they contribute to the factor $3zw$ in the front and the rest should only depend on $z^3$ and $w^3$. In fact the OPE of $W_{\frac{1}{3},\frac{\sqrt{2}}{9\sqrt{3}}}(z^3)V_{\frac{1}{\sqrt{3}}}(z^3)$ and $W_{\frac{1}{3},-\frac{\sqrt{2}}{9\sqrt{3}}}(w^3)V_{-\frac{1}{\sqrt{3}}}(w^3)$ is exactly $\frac{1}{z^3-w^3}$.

\subsection{A check for the insertion of $T$}

Recall that in \autoref{sec:naivebp} we found that, if we use the branched free-field construction
without considering the spectral network, the branch points naturally come with insertions of twist fields with conformal weight $\frac{1}{9}$.
Now we will give evidence in our simple example that using the nonabelianization map $\nab_\cW$ in \autoref{sec:nonabelianizationmap},
the behavior of the Virasoro generator $T$ near a branch 
point at $0$, i.e. $\IP{T(p)^z}_{\nab_\cW(\tPsi)}$ is regular. It is convenient to use the fermion 2-point function calculated in \autoref{tau:2point}.
As we did in \autoref{sec:naivebp}, we use 
a local coordinate $z= (x)^{\frac13}$, and we also simplify our notation a bit,
writing $\tJ(z)$ for $\tJ(p^{(1)})^{z^{(1)}}$, $\tJ(p^{(2)})^{z^{(2)}}$  or $\tJ(p^{(3)})^{z^{(3)}}$,
and similarly for $\psi_\pm$. 

Using the result of \cite{Hao:2024vlg}, we first compute:
\begin{equation} \label{eq:interm-3}
  \IP{\tT^\Heis(z) E_\ren(\cW) \cdots}_{\tPsi} = \lim_{w \to z} \frac{1}{2(z-w)} \left( K(z,w) - K(w,z) - \frac{2 \IP{E_\ren(\cW)\cdots}_{\tPsi}}{z-w} \right) \, .
\end{equation}
We can further express this in terms of the
fermion 2-point function with operator $E_\ren(\cW)$,
\begin{equation}
 K(z,y) = {\IP{\,
\psiplus(z)
\psiminus(y) E_\ren(\cW) \cdots}_\tPsi}\, .
\end{equation}
To be more precise: fix some open disc 
$U$ in $\bbC$, whose closure contains $z = 0$;
then thicken it to a small neighborhood of $U \times U$ in $\bbC^2$;
on this neighbourhood 
\begin{equation} \label{eq:K-approx}
   K(z,y) = 3 y z\left( \frac{\IP{E_\ren(\cW) \cdots}_\tPsi}{z^3-y^3} + a(z^3,y^3) \right)
\end{equation}
for a bounded function $a(x_1,x_2)$ such that $a(x,x)=0$. 
Substituting \eqref{eq:K-approx} in \eqref{eq:interm-3} gives
\begin{equation} \label{eq:interm-4}
{\IP{\tT^\Heis(z) E_\ren(\cW) \cdots}_{\tPsi}}= - \frac{1}{3z^2} {\IP{E_\ren(\cW)\cdots}_{\tPsi}} + O(z^4) \, .
\end{equation}
Now we are ready to compute the 
expectation value of $T^\tot$ in the conformal block
$\nab_\cW(\tPsi)$ on $C$. 
Recalling our dictionary \eqref{Ttot} reads $T^\tot(p)^z \rightsquigarrow \tT^\Heis(p^{(1)})^{z^{(1)}} + \tT^\Heis(p^{(2)})^{z^{(2)}}+ \tT^\Heis(p^{(3)})^{z^{(3)}}$. Using $z^{(n)} = \E^{\frac{2\pi\ri}{3}n}z^{(1)}$ (so $\de z^{(n)} / \de z^{(1)} = \E^{\frac{2\pi\ri}{3}n}$) and \eqref{eq:interm-4}, we get
\begin{equation}
{\IP{T^\tot(p)^z\cdots}_{\nab_\cW(\tPsi)}} = - \frac{1}{z^2} {\IP{\cdots}_{\nab_\cW(\tPsi)}}+ O(z^4) \, .
\end{equation}
This $- \frac{1}{ z^2}$ is just what we need to cancel the singularity coming from the change of coordinates. Indeed,
using 
the change-of-coordinate rule \eqref{eq:virasoro-coordinate-change}
at $c=3$ and $\{x,z\} = - \frac{4}{z^2}$, we get
\begin{equation}
  T^\tot(p)^{z} =(3z(p)^2)^2 T^\tot(p)^x + \frac{1}{4} \left(- \frac{4}{z(p)^2} \right) \, ,
\end{equation}
so
\begin{equation} \label{eq:T-branch-point-contributions-add}
  T^\tot(p)^x = \frac{1}{(3z(p)^2)^2} \left(-\frac{1}{ z(p)^2} +  \frac{1}{ z(p)^2} + O(z(p)^4)\right) = \regular\, ,
\end{equation}
as desired.

We should also check that there is no singularity in the Heisenberg correlators $\IP{J(p)^x \cdots}$ as $p$
approaches a branch point.
We directly compute:
\begin{align}
\IP{\tJ(z) E_\ren(\cW) \cdots}_\tPsi
&= \IP{\nop{\psiplus(z) \psiminus(z)} E_\ren(\cW) \cdots}_\tPsi \\
&=  \lim_{z' \to z} \left( \IP{\psiplus(z') \psiminus(z) E_\ren(\cW) \cdots} - \frac{\IP{E_\ren(\cW) \cdots}}{z'-z} \right)_\tPsi
\end{align}
and thus
\begin{equation}
 {\IP{\tJ(z) E_\ren(\cW) \cdots}_\tPsi}= \lim_{z' \to z} K(z',z) - \frac{{\IP{E_\ren(\cW) \cdots}_\tPsi} }{z'-z}.
\end{equation}
Substituting \eqref{eq:K-approx} here gives
\begin{equation}
\IP{\tJ(z) E_\ren(\cW) \cdots}_\tPsi = 0 \, .
\end{equation}
Then, using our dictionary 
 \eqref{eq:nab-operator-dictionary-J} we get
\begin{equation}
  {\IP{J(p)^z \cdots}_{\nab_\cW(\tPsi)}}=0\, ,
\end{equation}
and using \eqref{eq:heisenberg-coordinate-change} to change coordinates this implies
\begin{equation}
  J(p)^x = \frac{J(p)^z}{3z^2} = 0 \, .
\end{equation}

Now we can calculate $T = T^\tot - T^\Heis$ from the result of $T^\tot$ and $J$ above. It is straight forward to see that since $T^\tot$ and $J$ are both nonsingular at the branch point,
$T $ 
is also nonsingular there.

\subsection{$\tau$-function}

Now we are ready to look at the zero point function which is essential for an important conformal block know as the $\tau$-function. To be more precise,we calculate 
\begin{equation}\label{eq:0point}
\log \IP{E(\cW)}_{\tPsi}=\sum_{n=1}^\infty\sum_{\{\sigma_1,\cdots,\sigma_n\}}v_{\sigma_1\cdots \sigma_n},
\end{equation}
where
\begin{align}
\nonumber v_{\sigma_1\cdots \sigma_n}=&-\E^{-\frac{n\pi\ri}{2}+\frac{\pi}{2}\ri\sum\limits_{i=1}^n\sigma_i }\left(\frac{1}{2\pi\ri}\right)^n(\prod\limits_{i=1}^n\Theta_{(-\sigma_i)})\left(1+(\prod\limits_{i=1}^n\Theta_{(-\sigma_i)})\right)\left(1+(\prod\limits_{i=1}^n \Theta_{(-\sigma_i)})^2\right)\times\\
&\left(\prod\limits_{i=1}^n\,\int\limits_{\mathbb{R}_+}\rd t_i\right)  \left(\prod\limits_{i=2}^{n}\frac{1 }{\left( t_{i-1}-\e^{4 \ri \left(\alpha_{\sigma_i}-\alpha_{\sigma_{i-1}}\right)} \Theta_{{\sigma_i}}t_i\right)}\right)\frac{1 }{\left( t_n-\e^{4 \ri (\alpha_{\sigma_1}-\alpha_{\sigma_n})}\Theta_{{\sigma_1}}t_1\right)}.
\end{align}
As in \autoref{tau:2point}, we first transform 

\begin{equation}
f_0(t_1)= \frac{1}{\left( t_n-\e^{4 \ri (\alpha_{\sigma_1}-\alpha_{\sigma_n})} \Theta_{\sigma_1}t_1\right)}
\end{equation}
into
\begin{equation}
  \hat f_0(\lambda) = \frac{1}{2\pi}\int_{0}^\infty \phi_{-\lambda}(t_1) f_0(t_1) \, \de t_1 =\frac{1}2 \text{sech}(\pi\lambda)\E^{-\lambda\frac{\pi}{3}\sigma_1}\E^{\frac{\pi\ri}{6}\sigma_1}(\e^{4 \ri (\alpha_{\sigma_1}-\alpha_{\sigma_n})})^{-\frac12-\ri\lambda} \phi_{-\lambda}(t_n).
\end{equation}

Using the result \eqref{zwchecktheta}, we obtain

\begin{equation}
\begin{aligned}
v_{\sigma_1\cdots\sigma_n}
=&-\E^{-\frac{n\pi\ri}{2}+\frac{\pi}{2}\ri\sum\limits_{i=1}^n\sigma_i }\frac{1}{n}\left(1+(\prod\limits_{i=1}^n\Theta_{(-\sigma_i)})\right)\left(1+(\prod\limits_{i=1}^n \Theta_{(-\sigma_i)})^2\right)\left(\prod\limits_{i=1}^n\,\frac{1}{2\pi\ri}\int\limits_{\mathbb{R}_+}\rd t_i\right)\times\\
& \left(\prod\limits_{i=2}^{n}\frac{\Theta_{(-\sigma_i)} }{\left( t_{i-1}-\e^{4 \ri \left(\alpha_{\sigma_i}-\alpha_{\sigma_{i-1}}\right)} \Theta_{{\sigma_i}}t_i\right)}\right)\frac{\Theta_{(-\sigma_1)} }{\left( t_n-\e^{4 \ri (\alpha_{\sigma_1}-\alpha_{\sigma_n})} \Theta_{\sigma_1}t_1\right)}\\
=&-\left(\frac{1}{2\pi}\int\limits_{\mathbb{R}_+}\rd t\frac{1}{t}\right)\E^{\ri\pi n}\left(1+(\prod\limits_{i=1}^n\Theta_{\sigma_i})\right)\left(1+(\prod\limits_{i=1}^n \Theta_{\sigma_i})^2\right)\times\\
&\int_{-\infty}^\infty \frac{\left(\frac{1}{2}\sech(\pi\lambda)\right)^{n}}{n}\E^{-\ri\lambda(\sum\limits_{i=1}^n\log(-\Theta_{\sigma_i}))}\rd \lambda\\
=&-\E^{\ri\pi n}\left(\frac{1}{2\pi}\int\limits_{\mathbb{R}_+}\rd t\frac{1}{t}\right)\left(1+\E^{\frac{2\pi \ri}{3}(\sum\limits_{i=1}^n\sigma_i)}\right)\left(1+\E^{\frac{4\pi \ri}{3}(\sum\limits_{i=1}^n\sigma_i)}\right)\times\\
&\int_{-\infty}^\infty \frac{\left(\frac{1}{2}\sech(\pi\lambda)\right)^{n}}{n}\E^{-\frac\pi 3\lambda(\sum\limits_{i=1}^n\sigma_i)}\rd \lambda \, .
\end{aligned}
\end{equation}

To calculate \eqref{eq:0point}, we need to sum over the partition $\{\sigma_1,\cdots,\sigma_n\}$ for each $n$. 
Using \eqref{sumsigma} and \eqref{multsigma},
\begin{equation}\begin{aligned}\label{eq:log-divergent-result}
\log \IP{E(\cW)}_{\tPsi}&=\quad \sum_{n=1}^\infty\sum_{\{\sigma_1,\cdots,\sigma_n\}} v_{\sigma_1\cdots \sigma_n} \\
&=-\sum_{n=1}^\infty\sum_{k=0}^n \binom{n}{k}\left(\frac{1}{2\pi}\int\limits_{\mathbb{R}_+}\rd t_n\frac{1}{t_n} \right)\E^{\ri\pi n}\left(1+\E^{\frac{2\pi \ri}{3}(2k-n)}\right)\left(1+\E^{\frac{4\pi \ri}{3}(2k-n)}\right)\times\\
&\quad\quad\int_{-\infty}^\infty \frac{\left(\frac{1}{2}\sech(\pi\lambda)\right)^{n}}{n}\E^{-\frac\pi 3\lambda(2k-n)}\rd \lambda \\
&= \frac{1}{12}\int\limits_{\mathbb{R}_+}\rd t_n\frac{1}{t_n}=\frac{1}{3}\int\limits_{\mathbb{R}_+}\rd x_n\frac{1}{x_n}.
\end{aligned}
\end{equation}
This is exactly the scaling $\frac{1}{3}$ in \eqref{eq:renormalized-correlator}. To be more precise,
the result \eqref{eq:log-divergent-result} is logarithmically divergent, thus needs regularization.
We can cut off the integral near $0$ at $x_n = \eps_0$, and cut off the
integral near $\infty$ at $x_n = \eps_\infty^{-1}$ similarly. 
Then we get
\begin{equation}
  \log \IP{E(\cW)}_{\tPsi} = -\frac13 \log \eps_0 \eps_{\infty},
\end{equation}
and
\begin{equation}
  \IP{E(\cW)}_{\tPsi} = (\eps_0 \eps_\infty)^{-\frac13} \, .
\end{equation}
Note that this regularization is not exactly what we described
in our general scheme above; 
there we cut off all the integrals at a distance $\eps$,
rather than just the final one. However, the difference is reflected as the
normalization of the final result, but the form of the divergence is the same.
So we conclude that
\begin{equation}
  \IP{E(\cW)}_{\tPsi} = N (\eps_0 \eps_\infty)^{-\frac13} 
\end{equation}
for some normalization constant $N$. And this result justifies our claim in \autoref{sec:nonabelianizationmap}.

\section*{Acknowledgements}

We thank Pavlo Gavrylenko, Hans Jockers, Omar Kidwai, Pietro Longhi, Andrew Neitzke, Gerg\"{o} Nemes and Maximilian Schwick for helpful discussions.  QH’s work was supported by the Simons Foundation Collaboration Grant on New Structures in Low-dimensional
Topology (994320, JEA) and 
the Cluster of Excellence Precision Physics, Fundamental Interactions, and Structure of Matter (PRISMA++, EXC
2118) within the 
DFG - Project number:  390831469.

\appendix
\section{Two important identities in the calculation}
{\small
\begin{equation}\label{1wallsumtrick}
\int\limits_{\cG_{\alpha}}\frac{1}{(\tilde{z}-x)(x-\tilde{w}) }=\int\limits_{\mathbb{R}_+}\frac{ \E^{4 \ri \alpha}  (\tilde{w}+\tilde{z}) \left(\tilde{w}^2+\tilde{z}^2\right)}{\left(\tilde{z}^4-\E^{4 \ri
   \alpha} t\right)\left(\E^{4 \ri \alpha} t-\tilde{w}^4\right) }.
\end{equation}

\begin{equation}\label{1wallsumtrickgen}
\int\limits_{\cG_{\alpha}}\frac{\left(\tilde{z}+ \Theta_{\pm} x\right) \left(\tilde{z}^2+ (\Theta_{\pm}x)^2\right)}{(x-w) \left(\tilde{t}_1- (\Theta_{\pm} x)^4\right)}=\int\limits_{\mathbb{R}_+}\frac{ \e^{4 \ri \alpha }  \left(\tilde{z}+ \Theta_{\pm}  w\right) \left(\tilde{z}^2+ (\Theta_{\pm}w)^2\right)}{\left(\tilde{t}_1- \Theta_{\pm}t_2 e^{4 \ri
   \alpha  }\right)\left(e^{4 \ri \alpha } t_2-w^4\right) }=\int\limits_{\mathbb{R}_+}\frac{ \e^{4 \ri \alpha }  \left(\Theta_{\mp} \tilde{z}+w\right) \left(\left(\Theta_{\mp}  \tilde{z}\right)^2+w^2\right)}{\left(\tilde{t}_1-\Theta_{\pm} t_2 e^{4 \ri
   \alpha  }\right)\left(e^{4 \ri \alpha } t_2-w^4\right) }
\end{equation}
}
where $t_2=x^4$.

\section{Examples of terms in the expansion of $F(z,w)$}\label{appen:Fzw-exp}
\subsection{Example 1: 1-wall terms}

{\small
\begin{equation}
\begin{aligned}
w_+= &-\left(\frac{1}{2\pi\ri}\right)\e^{-\frac{2\pi\ri}{3}}\int\limits_{\cG_+} \frac{1}{z-x \E^{\frac{2\pi\ri}{3}}}\frac{1}{x-w}\rd x=-\left(\frac{1}{2\pi\ri}\right)\int\limits_{\mathbb{R}_+} \frac{ (-1)^{\frac23} \e^{4 \ri \alpha_+} \left(w+\Theta_- z\right) \left(w^2+(\Theta_- z)^2\right)}{\left( (\Theta_- z)^4-\e^{4 \ri \alpha_+} t\right)\left(\e^{4 \ri \alpha_+} t-w^4\right) }\rd t ,
 \end{aligned}
\end{equation}

\begin{equation}
\begin{aligned}
 w_-=&\left(\frac{1}{2\pi\ri}\right)\e^{\frac{2\pi\ri}{3}}\int_{\cG_{-}}\frac{1}{z-x \E^{-\frac{2\pi\ri}{3}}}\frac{1}{x-w}\rd x=\left(\frac{1}{2\pi\ri}\right)\int\limits_{\mathbb{R}_+} \frac{ (-1)^{-\frac23} \e^{4 \ri \alpha_-} \left(w+\Theta_+ z\right) \left(w^2+(\Theta_+ z)^2\right)}{\left( (\Theta_+ z)^4-\e^{4 \ri \alpha_-} t\right)\left(\e^{4 \ri \alpha_-} t-w^4\right) }\rd t.
 \end{aligned}
\end{equation}
}

\subsection{Example 2: 2-wall terms}

\begin{align}
\nonumber w_{++}=&\left(\frac{1}{2\pi\ri}\right)^2 \E^{-\frac{2\pi\ri}{3}\times 2}\int\limits_{\cG_+} \int\limits_{\cG_+} \frac{1}{z-x_1 \E^{\frac{2\pi\ri}{3}}}\frac{1}{x_1-x_2 \E^{\frac{2\pi\ri}{3}}}\frac{1}{x_2-w}\rd x\\
=&\left(\frac{1}{2\pi\ri}\right)^2 \int\limits_{\mathbb{R}_+} \int\limits_{\mathbb{R}_+} \frac{ \e^{4 \ri \alpha_+} }{\left( (\Theta_- z)^4-\e^{4 \ri \alpha_+} t_1\right)}\frac{\e^{4 \ri \alpha_+} \left(w+(\Theta_+z)\right)\left(w^2+(\Theta_+z)^2\right)}{\left(\e^{4 \ri \alpha_+} t_1-\e^{4 \ri \alpha_+} \Theta_+t_2\right)\left(\e^{4 \ri \alpha_+} t_2-w^4\right)}\rd t_1\rd t_2 ,\\
\nonumber w_{+-}=&-\left(\frac{1}{2\pi\ri}\right)^2 \int\limits_{\cG_+}\int\limits_{\cG_{-}} \frac{1}{z-x_1 \E^{\frac{2\pi\ri}{3}}}\frac{1}{x_1-x_2 \E^{-\frac{2\pi\ri}{3}}}\frac{1}{x_2-w}\rd x\\
=&-\left(\frac{1}{2\pi\ri}\right)^2 \E^{-\frac{2\pi\ri}{3}}\int\limits_{\mathbb{R}_+} \int\limits_{\mathbb{R}_+} \frac{ \e^{4 \ri \alpha_+} }{\left( (\Theta_- z)^4-\e^{4 \ri \alpha_+} t_1\right)}\frac{\e^{4 \ri \alpha_-} \left(w+z\right)\left(w^2+z^2\right)}{\left(\e^{4 \ri \alpha_+} t_1-\Theta_-\e^{4 \ri \alpha_-} t_2\right)\left(\e^{4 \ri \alpha_-  } t_2-w^4\right)}\rd t_1\rd t_2 ,\\
\nonumber w_{-+}=&-\left(\frac{1}{2\pi\ri}\right)^2 \int\limits_{\cG_{-}}\int\limits_{\cG_+} \frac{1}{z-x_1 \E^{\frac{-2\pi\ri}{3}}}\frac{1}{x_1-x_2 \E^{\frac{2\pi\ri}{3}}}\frac{1}{x_2-w}\rd x\\
=&-\left(\frac{1}{2\pi\ri}\right)^2 \e^{\frac{2\pi\ri}{3}}\int\limits_{\mathbb{R}_+} \int\limits_{\mathbb{R}_+} \frac{ \e^{4 \ri \alpha_-} }{\left( (\Theta_+ z)^4-\e^{4 \ri \alpha_-} t_1\right) }\frac{\e^{4 \ri \alpha_+}\left(w+z\right) \left(w^2+z^2\right)}{\left(\e^{4 \ri \alpha_-} t_1-\Theta_+\e^{4 \ri \alpha_+}t_2\right)\left(\e^{4 \ri \alpha_+}t_2-w^4\right)}\rd t_2\rd t_1,\\
\nonumber w_{--}=&\left(\frac{1}{2\pi\ri}\right)^2 \E^{\frac{2\pi\ri}{3}\times 2}\int\limits_{\cG_{-}}\int\limits_{\cG_{-}} \frac{1}{z-x_1 \E^{-\frac{2\pi\ri}{3}}}\frac{1}{x_1-x_2 \E^{-\frac{2\pi\ri}{3}}}\frac{1}{x_2-w}\rd x\\
=&\left(\frac{1}{2\pi\ri}\right)^2 \int\limits_{\mathbb{R}_+} \int\limits_{\mathbb{R}_+} \frac{ \e^{4 \ri \alpha_-} }{\left( (\Theta_+ z)^4-\e^{4 \ri \alpha_-} t_1\right)}\frac{\e^{4 \ri \alpha_-} \left(w+(\Theta_-z)\right)\left(w^2+(\Theta_-z)^2\right)}{\left(\e^{4 \ri \alpha_-} t_1-\e^{4 \ri \alpha_-} \Theta_-t_2\right)\left(\e^{4 \ri \alpha_-} t_2-w^4\right)}\rd t_1\rd t_2.
 \end{align}

\bibliography{w-algebra}
\end{document}